\documentclass[aps,prd,preprint,superscriptaddress,nofootinbib]{revtex4-1}
\usepackage{braket}
\usepackage{xargs}
\usepackage{mathtools}
\usepackage[english]{babel}
\usepackage[utf8]{inputenc}
\usepackage{amsmath,amssymb,braket,slashed,mathrsfs}
\usepackage{pifont}
\usepackage{MnSymbol}
\usepackage{graphicx}
\usepackage{tikz}
\usetikzlibrary{shapes,arrows,positioning}
\tikzset{decision/.style={diamond, draw, fill=blue!20, text width=4.5em, text badly centered, inner sep=0pt}}
\tikzset{block/.style={rectangle, draw, fill=blue!20, text width=10em, text centered, rounded corners, minimum width=3.5cm}}
\tikzset{block1/.style={rectangle, draw, fill=blue!20, text width=18.5em, text centered, rounded corners, minimum width=3.5cm}}
\tikzset{line/.style={draw, -latex, thick}}
\usepackage{color,hyperref}
\usepackage{multirow,array}

\usepackage{cleveref}

\newcommand{\ba}{\begin{eqnarray}}
\newcommand{\ea}{\end{eqnarray}}
\newcommand{\be}{\begin{equation}}
\newcommand{\ee}{\end{equation}}

\newcommand{\nn}{\nonumber}

\newcommand{\innovation}{Collaborative Innovation Center of Quantum Matter, Beijing 100871, China}
\newcommand{\chep}{Center for High Energy Physics, Peking University, Beijing 100871, China}
\newcommand{\pkuphy}{School of Physics, Peking University, Beijing 100871,
China}
\newcommand{\KeyLab}{State Key Laboratory of Nuclear Physics and Technology,
Peking University, Beijing 100871, China}
\newcommand{\Uconn}{Department of Physics, University of Connecticut, Storrs, CT 06269, USA}
\newcommand{\RBRC}{RIKEN-BNL Research Center, Brookhaven National Laboratory, Building 510, Upton, NY 11973}

\begin{document}
\title{Finite-volume formalism in the \boldmath $2 \xrightarrow[]{H_I+H_I} 2$ transition: an application
to the lattice QCD calculation of double beta decays}

\author{Xu~Feng}\email{xu.feng@pku.edu.cn}\affiliation{\pkuphy}\affiliation{\innovation}\affiliation{\chep}\affiliation{\KeyLab}
\author{Lu-Chang Jin}\email{ljin.luchang@gmail.com}\affiliation{\Uconn}\affiliation{\RBRC}
\author{Zi-Yu Wang}\affiliation{\pkuphy}
\author{Zheng Zhang}\affiliation{\pkuphy}
\pacs{PACS}

\date{\today}

\begin{abstract}

    We present the formalism for connecting a second-order electroweak $2\xrightarrow[]{H_I+H_I}2$ transition
    amplitudes in the finite volume (with two
    hadrons in the initial and final states) to the physical amplitudes in the infinite
    volume. Our study
    mainly focus on the
    case where the low-lying intermediate state consists of
    two scattering hadrons. As a side product we also reproduce the finite-volume
    formula for $2\xrightarrow[]{H_I}2$ transition, originally obtained 
    by Brice\~no and Hansen in Ref.~\cite{Briceno:2015tza}.
    With the available finite-volume formalism, we further
    discuss how to treat with the finite-volume problem in the double beta
    decays $nn\to pp ee\bar{\nu}\bar{\nu}$ and $nn\to pp ee$. 

\end{abstract}

\maketitle

\section{Introduction}

Lattice QCD provides a well-established non-perturbative approach to 
solve the quantum chromodynamics (QCD) theory of quarks and gluons. Using the
high-performance supercomputers, the quarks and gluons are enclosed and simulated in a discretized,
finite-volume lattice. Controlling the various systematic effects such as lattice
discretization effects, finite-volume effects, and unphysical quark mass effects
  is required for lattice QCD calculation to make the high-precision predication from
first principles. On the other hand, in some cases the study of the
systematic effects is much more than the reduction of the uncertainty. It
could lead to the new methodology to solve the interesting physics problems. For
example, the study of the pion mass dependence from lattice QCD interplays with the chiral
perturbation theory, yielding a deeper understanding of the chiral dynamics of
QCD. Another example is the pioneering work on the finite-volume formalism by
L\"uscher~\cite{Luscher:1986pf,Luscher:1990ux,Luscher:1991cf}.
It allows us to connect the discrete energy spectrum calculated from lattice QCD
to the infinite-volume scattering phase and has played an important role in
understanding the hadron spectra and hadron-hadron scattering.

The finite-volume formalism generically includes three topics.
\begin{itemize}
\item Finite-volume energy quantization relates the discrete energy in the
    finite volume to the scattering phase in the infinite volume. 
        The best examples under well investigation are the pion-pion scattering 
        in the isospin
        $I=2$~\cite{Yamazaki:2004qb,Beane:2005rj,Beane:2007xs,Feng:2009ij,Dudek:2010ew,Beane:2011sc,Fu:2013ffa,Sasaki:2013vxa,Dudek:2012gj,Helmes:2015gla,Culver:2019qtx}, $I=1$
        ($\rho$ resonance
        relevant)~\cite{Aoki:2007rd,Feng:2010es,Lang:2011mn,Aoki:2011yj,Pelissier:2012pi,Dudek:2012xn,Feng:2014gba,Wilson:2015dqa,Bali:2015gji,Bulava:2016mks,Guo:2016zos,Fu:2016itp,Alexandrou:2017mpi,Andersen:2018mau,Werner:2019hxc,Erben:2019nmx},
        and $I=0$ ($\sigma$ resonance and disconnected diagrams
        relevant)~\cite{Liu:2009uw,Briceno:2016mjc,Liu:2016cba,Briceno:2017qmb,Fu:2017apw,Guo:2018zss,Wang:2019nes}
        channels. Due to the good signals provided by the pion-pion system, a lot of
        attentions are paid to these scattering channels in the past years.
        For more lattice calculations of scattering amplitudes, we
        refer to a recent review~\cite{Briceno:2017max}.
    \item Lellouch-L\"uscher relation~\cite{Lellouch:2000pv} connects the finite-volume matrix element with
    two hadrons in either initial or final state to the physical matrix element
        in the
        infinite volume. Such examples include $0\xrightarrow[]{J}2$ decays e.g. 
        the timelike pion form
        factor~\cite{Meyer:2011um,Feng:2014gba,Andersen:2018mau,Erben:2019nmx},
        $1\xrightarrow[]{J}2$ decays
        including
        $K\to\pi\pi$~\cite{Blum:2011ng,Blum:2012uk,Blum:2015ywa,Bai:2015nea,Abbott:2020hxn} and $\pi\pi\to\pi\gamma^*$
        transition~\cite{Briceno:2015dca,Briceno:2016kkp,Alexandrou:2018jbt} and
        $2\xrightarrow[]{J}2$ decays recently studied in
        Refs.~\cite{Briceno:2015tza,Baroni:2018iau}.
    \item Finite-volume formula for long-distance electroweak
        amplitudes~\cite{Christ:2010gi,Christ:2014qaa,Christ:2015pwa} relates the
        bilocal matrix element in the finite volume to the physical one in the
        infinite volume. This formalism is first developed to solve the
        finite-volume problem for $K_L$-$K_S$
        mixing~\cite{Christ:2012se,Bai:2014cva,Bai:2016gzv,Wang:2020jpi} and has
        been used for other
        second-order electroweak processes such as rare kaon decays~\cite{Christ:2015aha,Christ:2016eae,Christ:2016mmq,Bai:2017fkh,Bai:2018hqu,Christ:2019dxu}. Recently
        the formalism is generalized in Ref.~\cite{Briceno:2019opb} to access more long-distance
        observables.
\end{itemize}
It is found by Ref.~\cite{Christ:2015pwa} that the above three finite-volume
formulae can be derived in a uniform way in the framework of quantum field
theory using the techniques of Kim,
Sachrajda and Sharpe (KSS)~\cite{Kim:2005gf}. 

In this work, we present the derivation of the finite-volume formula for
long-distance electroweak amplitudes with two hadrons in both initial and final
states ($2\xrightarrow[]{H_I+H_I}2$).
We consider the scattering process with two channels, which are mixed by the electroweak interaction. 
We label these two channels by
$\alpha$ and $\beta$.
The master formula is given as
\be
\label{eq:correction_formula}
\frac{d\left(\phi+\delta_\alpha^{(0)}\right)}{dE}\Delta
E_\alpha+\Delta\delta_\alpha=\frac{1}{4}\cot\left(\phi+\delta_\beta^{(0)}\right)|\langle
E,\mathrm{in},\beta|H_I|E,\mathrm{in},\alpha\rangle|^2,\quad\mbox{at
$E=E_\alpha^{(0)}$},
\ee
where $E_\alpha^{(0)}$ is discrete energy for initial/final state without
non-QCD correction. $\Delta E_\alpha$ is the energy shift when
turning on the second-order electroweak interaction, and it equals to the 
$2\xrightarrow[]{H_I+H_I}2$ finite-volume matrix element calculated on the
lattice. $\phi$ is a known, kinematic function,
originally introduced by L\"uscher in Eq.~(6.12) of Ref.~\cite{Luscher:1990ux}. 
$\delta_\alpha^{(0)}$ is the strong scattering phase for the
initial/final state and $\delta_\beta^{(0)}$ is the scattering phase for the
low-lying two-hadron intermediate state. Here we consider the case that the
lowest intermediate state consists of two interacting hadrons.
$\Delta\delta_\alpha$ is the shift in the total scattering phase with the
existence of non-QCD interaction. It is equivalent to the infinite-volume 
$2\xrightarrow[]{H_I+H_I}2$
matrix element as we explain later.
The derivation is performed using the perturbative approach proposed by Lellouch
and L\"uscher~\cite{Lellouch:2000pv} together with the coupled-channel finite-volume energy 
quantization condition~\cite{He:2005ey,Liu:2005kr}.
As a side product, we also obtain the finite-volume formula for 
$2\xrightarrow[]{J}2$ transition for the special case that the current $J$
carries the vanishing momentum. For more general cases, we refer to
Refs.~\cite{Briceno:2015tza,Baroni:2018iau}.

We find that 
the KSS approach~\cite{Kim:2005gf} treats the finite-volume problem in a
thorough and fundamental
way using Poisson summation formula. Many new developments of the
finite-volume formalism are made progress along the direction proposed by KSS.
On the other hand, the approach
invented by Lellouch and L\"uscher~\cite{Lellouch:2000pv} creates another possibility that one can
obtain the finite-volume formalism in a relatively simple way as
the finite-volume information is already incorporated inside
L\"uscher's quantization condition and it is not necessary to investigate it
again using Poisson summation formula.

The paper is organized as follows. In Sect.~\ref{sect:finite_volume}, we discuss
the discrete energy shift in the finite volume due to the existence of the $2 \xrightarrow[]{H_I+H_I} 2$ transition. 
In Sect.~\ref{sect:infinite_volume}, we discuss the infinite-volume scattering
amplitude relevant for the $2 \xrightarrow[]{H_I+H_I} 2$
transition. In Sect.~\ref{sect:correction_formula}, the energy shift is related to the scattering
amplitude using the coupled-channel quantization condition and thus the
finite-volume formalism~Eq.~(\ref{eq:correction_formula}) is obtained. In
Sect.~\ref{sect:application}, we discuss the
applications of the finite-volume formalism to the double beta decays.

\section{\boldmath $2 \xrightarrow[]{H_I+H_I} 2$ transition in the finite volume}
\label{sect:finite_volume}

We consider the full Hamiltonian including both QCD and non-QCD interactions as
\be
H^L=H_0^{L}+H_I^{L},
\ee
where $H_0^{L}$ stands for the pure strong interaction and $H_I^{L}$ indicates
the non-QCD ones, e.g. electromagnetic or weak interactions.
The superscript $L$ reminds us that all the interactions are
constrained by a finite volume.

When the interaction $H_I$ is turned on, it is possible that two independent
strong scattering (or bound) channels are mixed by the non-QCD interaction. For example, in the
double beta decay, the ${^1S_0}$ two-nucleon state can mix with the ${^3S_1}$
state. To specify this character of the $2\xrightarrow[]{H_I + H_I} 2$ transition, we
assign two low-lying eigenstates of the Hamiltonian $H_0^{L}$ as $|\alpha\rangle^{L}$
and
$|\beta\rangle^{L}$, which satisfy the normalization conditions 
\be
{^{L}}\langle \alpha|H_0^{L}|\alpha\rangle^{L}=E_\alpha^{(0)},\quad
{^{L}}\langle \beta|H_0^{L}|\beta\rangle^{L}=E_\beta^{(0)},\quad
{^{L}}\langle \beta|H_0^{L}|\alpha\rangle^{L}=0,
\ee
and $E_\alpha^{(0)}$ and $E_\beta^{(0)}$ are the corresponding energy
eigenvalues.
These two states are independent when turning off the non-QCD interactions but
mix with each other when turning on these interactions.
In the finite volume, the spectra of QCD Hamiltonian is
discrete and it allows for multiple nearly-degenerate states. Here we
focus on only one of them and classify all the other states as
$|n_\alpha\rangle^{L}$
and $|n_\beta\rangle^{L}$, where $|n_\alpha\rangle^{L}$ and
$|n_\beta\rangle^{L}$ have
the same quantum number as
$|\alpha\rangle^{L}$ and $|\beta\rangle^{L}$, respectively.
We introduce the projectors
\be
Q=\sum_{n=\alpha,\beta}|n\rangle^{L}{^{L}}\langle n|,\quad
P=1-Q,
\ee
to construct a two-state subspace.

The eigenvalue equation for the full Hamiltonian is given by
\be
(H_0^L+H_I^L)|n\rangle_I^L=E_n|n\rangle_I^L.
\ee
In the notation of the eigenstate $|n\rangle_I^L$ the subscript $I$ is used to
indicate the existence of the non-QCD interaction.
Acting $P$ and $Q$ on the above equation, we have
\ba
&&H_0^LP|n\rangle_I^L+PH_I^L(Q+P)|n\rangle_I^L=E_nP|n\rangle_I^L,
\nn\\
&&H_0^LQ|n\rangle_I^L+QH_I^L(Q+P)|n\rangle_I^L=E_nQ|n\rangle_I^L.
\ea
This results in
\ba
\label{eq:projection}
&& P H_I^LQ|n\rangle^L_I=(E_n-H_0^L-P H_I^L P)P|n\rangle^L_I,
\nn\\
&& Q H_I^LP|n\rangle^L_I=(E_n-H_0^L-Q H_I^L Q)Q|n\rangle^L_I.
\ea
Inserting $P|n\rangle^L_I=P(E_n-H_0^L-P H_I^L P)^{-1}P H_I^LQ|n\rangle_I^L$
into the second line of Eq.~(\ref{eq:projection}), we have
\be
QH_I^LP(E_n-H_0^L-PH_I^LP)^{-1}PH_I^LQ|n\rangle^L_I
=(E_n-H_0^L-Q H_I^L Q)Q|n\rangle^L_I.
\ee
By neglecting the $O(H_I^3)$ terms, we obtain the equations
\be
\left(\tilde{H}_0+\tilde{H}_I\right)Q|n\rangle^L_I=E_n\,Q|n\rangle^L_I,
\ee
with
\be
\tilde{H}_0=H_0^L+Q H_I^L Q,\quad \tilde{H}_I=QH_I^LP(E_n-H_0^L)^{-1}PH_I^LQ.
\ee
The existence of the nonzero solutions for equations
\ba
&&{^L}\langle\alpha|\tilde{H}_0+\tilde{H}_I|\alpha\rangle^L{^L}\langle\alpha|\alpha\rangle_I^L+{^L}\langle\alpha|\tilde{H}_0+\tilde{H}_I|\beta\rangle^L{^L}\langle\beta|\alpha\rangle_I=E_\alpha\,{^L}\langle\alpha|\alpha\rangle_I^L,
\nn\\
&&{^L}\langle\beta|\tilde{H}_0+\tilde{H}_I|\alpha\rangle^L{^L}\langle\alpha|\alpha\rangle_I^L+{^L}\langle\beta|\tilde{H}_0+\tilde{H}_I|\beta\rangle^L{^L}\langle\beta|\alpha\rangle_I=E_\alpha\,{^L}\langle\beta|\alpha\rangle_I^L,
\ea
requires that the secular equation holds
\be
\begin{vmatrix}
    {^L}\langle\alpha|\tilde{H}_0+\tilde{H}_I|\alpha\rangle^L-E_\alpha &
    {^L}\langle\alpha|\tilde{H}_0+\tilde{H}_I|\beta\rangle^L \\
    {^L}\langle\beta|\tilde{H}_0+\tilde{H}_I|\alpha\rangle^L &
    {^L}\langle\beta|\tilde{H}_0+\tilde{H}_I|\beta\rangle^L-E_\alpha 
\end{vmatrix}=0.
\ee
For the general case with $E_\alpha^{(0)}\neq E_\beta^{(0)}$, the solution of $E_\alpha$
is given by
\be
\label{eq:Delta_E}
E_\alpha=E_\alpha^{(0)}+\Delta E_\alpha,\quad
\Delta
E_\alpha=\frac{\left|{^L}\langle\beta|H_I^L|\alpha\rangle^L\right|^2}{E_\alpha^{(0)}-E_{\beta}^{(0)}}+\sum_{n_\beta\neq\beta}\frac{\left|{^L}\langle
n_\beta|H_I^L|\alpha\rangle^L\right|^2}{E_\alpha^{(0)}-E_{n_\beta}^{(0)}}.
\ee
The energy shift $\Delta E_\alpha$ is exactly the finite-volume long-distance matrix element obtained
from a lattice QCD calculation.

Here we obtain Eq.~(\ref{eq:Delta_E}) using the second-order degenerate perturbation theory.
In fact Eq.~(\ref{eq:Delta_E}) is the standard result from the second-order perturbation
theory and we expect the derivation could be simpler using the common perturbation
theory.

\section{\boldmath $2 \xrightarrow[]{H_I + H_I} 2$ transition in the infinite volume}
\label{sect:infinite_volume}

Now we consider the $2\xrightarrow[]{H_I + H_I}2$ transition in the infinite volume. For
simplicity we only discuss the case that the low-lying intermediate state is given by a two-particle scattering state
or a one-particle bound state. For the former, the transition amplitude involves
the input of a $2\times2$ scattering $S$-matrix. For the latter, a
single-channel $S$-matrix is relevant.

\subsection{Process of \boldmath $2 \xrightarrow[]{H_I}  2 \xrightarrow[]{H_I} 2$}

We first consider the scattering state by turning off the non-QCD interactions. 
In the infinite volume, we use $|E,\mathrm{in},\alpha\rangle$ to describe the incoming scattering state
and $\langle E,\mathrm{out},\alpha|$ for the outgoing scattering state. The
low-lying intermediate scattering state is described by $|E,\mathrm{in},\beta\rangle$. 
For simplicity, here we only consider the S-wave scattering. The relevant
normalization condition is assigned as
\be
\langle
E',\mathrm{in},\beta|E,\mathrm{in},\alpha\rangle=2\pi\delta(E-E')\delta_{\alpha\beta}.
\ee
The scattering $S$-matrix is defined as
\be
\begin{pmatrix}
    \langle E',\mathrm{out},\alpha|E,\mathrm{in},\alpha\rangle & \langle
    E',\mathrm{out},\beta|E,\mathrm{in},\alpha\rangle \\
    \langle E',\mathrm{out},\alpha|E,\mathrm{in},\beta\rangle & \langle
    E',\mathrm{out},\beta|E,\mathrm{in},\beta\rangle
\end{pmatrix}
=2\pi\delta(E-E')S,\quad
S=
\begin{pmatrix}
    e^{2i\delta_\alpha^{(0)}} & 0 \\
    0 & e^{2i\delta_\beta^{(0)}}
\end{pmatrix}.
\ee
Without non-QCD interactions, there is no mixing between $\alpha$ and $\beta$
states. Thus $S$ is a diagonal matrix with $\delta_\alpha^{(0)}$ and
$\delta_\beta^{(0)}$ the scattering phases for pure strong interaction. 
We use $|\alpha'\rangle$ and $|\beta'\rangle$ to stand for the excited states,
which have the same quantum number as $|E,\mathrm{in},\alpha\rangle$ and
$|E,\mathrm{in},\beta\rangle$, respectively. 
We assume that the threshold energy $E_{\mathrm{th}}$ for these excited states are above the
energy region we are interested in.

When turning on the non-QCD interactions, the scattering state for full
Hamiltonian $H=H_0+H_I$ is given by
\be
\label{eq:scattering_state}
|E,\mathrm{in},\alpha\rangle_I=|E,\mathrm{in},\alpha\rangle+G_E^{(+)}H_I|E,\mathrm{in},\alpha\rangle_I,
\ee
where
\be
G_E^{(+)}=\frac{1}{E-H_0+i\varepsilon}=\mathcal{PV}\frac{1}{E-H_0}-i\pi\delta(E-H_0)
\ee
is the standard Green's function. With non-QCD interactions, we parameterize the
$S$-matrix following Refs.~\cite{He:2005ey,Liu:2005kr}
\be
S_I=
\begin{pmatrix}
c\,e^{2i\delta_\alpha} & i\,s\,e^{i\delta_\alpha+i\delta_\beta} \\
i\,s\,e^{i\delta_\alpha+i\delta_\beta} & c\,e^{2i\delta_\beta} 
\end{pmatrix},
\ee
where the real values $c$ and $s$ satisfy the relation $c^2+s^2=1$.
This parameterization makes the derivation of the finite-volume formalism
very straightforward.
(In some other cases, e.g. in the $K\to\pi\pi$
decay where $I=0$ and $I=2$ $\pi\pi$ states mix due to the existence of
electromagnetic interactions~\cite{Christ:2017pze}, it is simpler to use the parameterization proposed by
Ref.~\cite{Hansen:2012tf}.)

It is useful to relate the $S$-matrix to the $T$-matrix using the relation $S=1+iT$. After
turning on the non-QCD interaction, the change of the $T$ matrix is given by
\be
\label{eq:Delta_T1}
\Delta T=-i
\begin{pmatrix}
    c\,e^{2i\delta_\alpha}-e^{2i\delta_\alpha^{(0)}} & i\,s\,e^{i\delta_\alpha+i\delta_\beta} \\
i\,s\,e^{i\delta_\alpha+i\delta_\beta} & c\,e^{2i\delta_\beta}
    -e^{2i\delta_\beta^{(0)}}
\end{pmatrix}.
\ee
On the other hand, the matrix of $\Delta T$ can be constructed using the scattering
state through
\be
\Delta T=
-\begin{pmatrix}
    \langle E,\mathrm{out},\alpha|H_I|E,\mathrm{in},\alpha\rangle_I 
    & \langle E,\mathrm{out},\beta|H_I|E,\mathrm{in},\alpha\rangle_I \\
    \langle E,\mathrm{out},\alpha|H_I|E,\mathrm{in},\beta\rangle_I & \langle
    E,\mathrm{out},\beta|H_I|E,\mathrm{in},\beta\rangle_I \\
\end{pmatrix}.
\ee
We can make the perturbative expansion of $\Delta T$. Up to $O(H_I^2)$, we find
\be
\label{eq:Delta_T2}
\Delta T=-\begin{pmatrix}
    e^{2i\delta_\alpha^{(0)}}
    (K_\alpha-i|J|^2/2)
    & e^{i\delta_\alpha^{(0)}+i\delta_\beta^{(0)}}J \\
    e^{i\delta_\alpha^{(0)}+i\delta_\beta^{(0)}} J^* &
    e^{2i\delta_\beta^{(0)}}(K_\beta-i|J|^2/2) \\
\end{pmatrix},
\ee
where
\ba
K_\alpha&=&\mathcal{PV}\int\frac{dE'}{2\pi}\,\frac{
|\langle
E',\mathrm{in},\beta|H_I|E,\mathrm{in},\alpha\rangle|^2}{E-E'}
+\sumint_{\beta'}
\frac{
|\langle
\beta'|H_I|E,\mathrm{in},\alpha\rangle|^2}{E-E_{\beta'}},
\nn\\
J&=&e^{i\delta_\beta^{(0)}-i\delta_\alpha^{(0)}}\langle
E,\mathrm{in},\beta|H_I|E,\mathrm{in},\alpha\rangle.
\ea
Here we have used the simplified symbol
$\sumint_{\beta'}\equiv\sum_{\beta'}\int_{E_{\mathrm{th}}}^{\infty}\frac{dE_{\beta'}}{2\pi}$.
Under the symmetry of the time reversal invariance, $J$ is a real quantity.
By exchanging $\alpha$ and $\beta$ for $K_\alpha$, one gets the expression for
$K_\beta$.

Equating Eqs.~(\ref{eq:Delta_T1}) and (\ref{eq:Delta_T2}), we obtain
\be
\label{eq:Delta_delta}
s=-J,\quad \Delta\delta_\alpha\equiv \delta_\alpha-\delta_\alpha^{(0)}=-\frac{K_\alpha}{2},\quad
\Delta\delta_\beta\equiv\delta_\beta-\delta_\beta^{(0)}=-\frac{K_\beta}{2}.
\ee

\subsection{Process of \boldmath $2\xrightarrow[]{H_I} 1 \xrightarrow[]{H_I} 2$}

For the $2\xrightarrow{H_I+H_I}2$ process with a deeply bound intermediate state, the
first example comes from $\pi\pi\to K\to\pi\pi$ in L. Lellouch and
M. L\"uscher's work~\cite{Lellouch:2000pv}.
Later, H. Meyer extended it to the case of $\pi\pi\to W\to\pi\pi$~\cite{Meyer:2011um},
where a massive gauge boson $W$ is introduced and annihilate with an auxiliary
vector field to obtain a
finite-volume formula for the timelike pion form factor.
In Ref.~\cite{Christ:2010gi}, N. Christ used again the $\pi\pi\to K\to\pi\pi$ transition
amplitude to obtain a finite-volume correction for the $K_L$-$K_S$ mass
difference. Here we include the process of $2 \xrightarrow[]{H_I}  1
\xrightarrow[]{H_I} 2$ 
simply for the completeness of the discussion.

If $\beta$ is a deeply bound state, it is not necessary to introduce a $2\times2$
$S$-matrix. The correction to the $T$-matrix due to the non-QCD interaction is
given by
\be
\Delta T= -\langle E,\mathrm{out},\alpha|H_I|E,\mathrm{in},\alpha\rangle_I.
\ee
Using Eq.~(\ref{eq:scattering_state}) and inserting the $|\beta\rangle$ and $|\beta'\rangle$ states into $\Delta T$ 
one can obtain
\be
\label{eq:Delta_T_bound}
\Delta
T=-e^{2i\delta_\alpha^{(0)}}\left(\frac{|\langle\beta|H_I|E,\mathrm{in},\alpha\rangle|^2}{E-E_\beta}
+\sumint_{\beta'}\frac{|\langle\beta'|H_I|E,\mathrm{in},\alpha\rangle|^2}{E-E_{\beta'}}\right).
\ee
It results in
\be
\label{eq:Delta_delta_bound}
\Delta\delta_\alpha\equiv\delta_\alpha-\delta_\alpha^{(0)}=-\frac{\hat{K}_\alpha}{2},\quad
\hat{K}_\alpha=\frac{|\langle\beta|H_I|E,\mathrm{in},\alpha\rangle|^2}{E-E_\beta}
+\sumint_{\beta'}\frac{|\langle\beta'|H_I|E,\mathrm{in},\alpha\rangle|^2}{E-E_{\beta'}}.
\ee

\section{Finite-volume formalism}
\label{sect:correction_formula}

In this section we present the finite-volume formalism which connects the
matrix elements that can be calculated in the finite volume using lattice QCD to
the infinite-volume transition amplitudes. 

We first discuss the $2\xrightarrow[]{H_I}2\xrightarrow[]{H_I}2$ transition. The coupled-channel finite-volume
energy quantization condition has been first established by
Refs.~\cite{He:2005ey,Liu:2005kr} in 2005 using the quantum
mechanics. 
Later, there have been a number of papers studying the generalization
of L\"uscher's quantization condition to multiple
channels~\cite{Lage:2009zv,Bernard:2010fp,Doring:2011vk,Aoki:2011gt,Hansen:2012tf}. For
example, in Ref.~\cite{Hansen:2012tf} quantization condition is extended to quantum field
theory using the KSS approach~\cite{Kim:2005gf}.

When turning on the non-QCD interaction, we adopt the quantization condition from Refs.~\cite{He:2005ey,Liu:2005kr}
\be
\label{eq:coupled_channel}
\left(e^{-2i(\phi+\delta_\alpha)}-c\right)\left(e^{-2i(\phi+\delta_\beta)}-c\right)+s^2=0,\quad\mbox{at
$E=E_\alpha$},
\ee
where the angle $\phi$ is a known function of discrete, finite-volume energy
$E$~\cite{Luscher:1990ux}. 
(By multiplying a factor of $e^{2i\delta_\alpha+2i\delta_\beta}$, 
Eq.~(\ref{eq:coupled_channel}) can reproduce Eq.~(34) in Ref.~\cite{He:2005ey}.)
When turning off the non-QCD interaction we have
\be
\label{eq:single_channel}
e^{-2i\left(\phi+\delta_\alpha^{(0)}\right)}-1=0,\quad\mbox{at $E=E_\alpha^{(0)}$}.
\ee
Comparing Eqs.~(\ref{eq:coupled_channel}) and (\ref{eq:single_channel}) and
using the relation $s^2=|\langle
E,\mathrm{in},\beta|H_I|E,\mathrm{in},\alpha\rangle|^2$ given in
Eq.~(\ref{eq:Delta_delta}), we
obtain the master formula given in Eq.~(\ref{eq:correction_formula}). We copy it
here for the sake of an easier read
\be
\label{eq:correction_formula_copy}
\frac{d\left(\phi+\delta_\alpha^{(0)}\right)}{dE}\Delta
E_\alpha+\Delta\delta_\alpha=\frac{1}{4}\cot\left(\phi+\delta_\beta^{(0)}\right)|\langle
E,\mathrm{in},\beta|H_I|E,\mathrm{in},\alpha\rangle|^2,\quad\mbox{at
$E=E_\alpha^{(0)}$},
\ee
where $\Delta E_\alpha$ is the finite-volume matrix element defined in
Eq.~(\ref{eq:Delta_E}) and $\Delta \delta_\alpha$ is the infinite-volume matrix element
defined in Eq.~(\ref{eq:Delta_delta}). It is not surprising that the finite-volume correction
formula takes the form of Eq.~(\ref{eq:correction_formula_copy}) as the initial/final state receives a
correction of Lellouch-L\"uscher factor $\frac{d\left(\phi+\delta_\alpha^{(0)}\right)}{dE}$
and the intermediate state receives a correction of factor
$\cot\left(\phi+\delta_\beta^{(0)}\right)$ as first obtained by
Refs.~\cite{Christ:2014qaa,Christ:2015pwa}. It is known that the energy
quantization condition can be used for a shallow bound state through the
analytical continuation~\cite{Beane:2003da,Sasaki:2006jn}. Thus the master formula derived here can be extended
from a scattering state to a shallow bound state.

In the limit of $E_\beta^{(0)}\to E_\alpha^{(0)}$, both $\Delta E_\alpha$ and
$\cot\left(\phi+\delta_\beta^{(0)}\right)$ in
Eq.~(\ref{eq:correction_formula_copy})
become singular. By equating the residue of the poles, we obtain
\be
\label{eq:2to2}
h_\alpha'\left|{^L}\langle\beta|H_I^L|\alpha\rangle^L\right|^2
h_\beta'=\frac{1}{4}\left|\langle
E,\mathrm{in},\beta|H_I|E,\mathrm{in},\alpha\rangle\right|^2,\quad\mbox{at
$E=E_\alpha^{(0)}$ and
$E_\beta^{(0)}\to E_\alpha^{(0)}$},
\ee
where $h_{i}=\phi+\delta_{i}^{(0)}$ and
$h_{i}'=dh_{i}/dE$ for $i=\alpha,\beta$.
We thus reproduce the finite-volume correction formula for the $2\xrightarrow[]{J}2$
transition matrix with the current $J$ carrying zero momentum, which is first obtained by Ref.~\cite{Briceno:2015tza}.

For the $2\xrightarrow[]{H_I}1\xrightarrow[]{H_I}2$ transition, the corresponding finite-volume formula
is given by
\be
\label{eq:correction_formula1}
\frac{d\left(\phi+\delta_\alpha^{(0)}\right)}{dE}\Delta
E_\alpha+\Delta\delta_\alpha=0,
\ee
where $\Delta E_\alpha$ is given by Eq.~(\ref{eq:Delta_E}) and $\Delta\delta_\alpha$ is
given by Eq.~(\ref{eq:Delta_delta_bound}).

\section{Application to double beta decays}
\label{sect:application}

Observation of neutrinoless double beta ($0\nu2\beta$) decays would prove neutrinos as 
Majorana fermions and lepton number violation in nature. As a result the study of
double beta decays attracts a lot of interests from both experimental and
theoretical sides.
Current knowledge of second-order weak-interaction nuclear matrix elements needs to be improved, 
as various nuclear models lead to discrepancies on the order of 100\%~\cite{Engel:2016xgb}. 
A promising approach to improving the reliability of the theoretical predication is to
combine the chiral effective field theory
($\chi$EFT)~\cite{Cirigliano:2017ymo,Cirigliano:2017djv,Cirigliano:2017tvr,Pastore:2017ofx,Cirigliano:2018hja,Cirigliano:2018yza,Cirigliano:2019vdj,Dekens:2020ttz} with lattice QCD and
then provide
well-constrained few-body inputs to ab initio many-body calculations~\cite{Engel:2016xgb}. 
Efforts have been invested to calculate double beta decays in both 
pion~\cite{Nicholson:2018mwc,Feng:2018pdq,Detmold:2018zan,Tuo:2019bue,Detmold:2020jqv} and
nucleon~\cite{Tiburzi:2017iux,Shanahan:2017bgi} sector from lattice QCD.

We start the discussion of the finite-volume problem for the double beta decays in the pion sector, taking the
$\pi^-\pi^-\to\pi^- e\nu\to ee$ and $\pi^-\to \pi^0e\nu\to \pi^+ee$ as examples. If we only
consider the hadronic particles, the former
process is a $2\xrightarrow[]{H_I}1\xrightarrow[]{H_I} 0$ transition and the
latter is a $1\xrightarrow[]{H_I}
1\xrightarrow[]{H_I}1$ transition. However, one
needs to pay attention to
the finite-volume effects
caused by the massless neutrino in the intermediate state. For the case of
$\pi^-\pi^-\to\pi^- e\nu\to ee$ transition, there are two
sources of power-law
finite-volume effects~\cite{Feng:2018pdq}. One arises from the $\pi^-\pi^-$ initial state
and is corrected by the inclusion of Lellouch-L\"uscher factor. The other one
originates from the massless neutrino and is estimated as an $O(L^{-2})$ effect by using
the QED$_{\mathrm{L}}$ technique. In the study of $\pi^-\to \pi^0e\nu\to \pi^+ee$ transition~\cite{Tuo:2019bue}, 
a novel method called
{\em infinite-volume reconstruction}~\cite{Feng:2018qpx} is used to treat the massless
neutrino in the intermediate state. This method
reduces the usual power-law finite-volume effect induced by the neutrino-pion loop to an
exponentially suppressed effect. With the finite-volume
corrections, Refs.~\cite{Feng:2018pdq,Tuo:2019bue} produce the lattice results for the double beta decay
amplitudes, which
are well consistent with
the $\chi$EFT formula~\cite{Cirigliano:2017tvr} and 
much more accurate than the estimates from the phenomenological
study~\cite{Ananthanarayan:2004qk}. In an exploratory study~\cite{Detmold:2018zan},
Detmold and Murphy make an attempt to use massive
neutrino for $\pi^-\to \pi^0e\nu\to \pi^+ee$
and then study the neutrino mass
dependence. (In a recent work~\cite{Detmold:2020jqv}, the authors use the massless
neutrinos in their latest results, where power-law finite-volume effect is a
relevant issue.) 
We consider the massive neutrino a good solution to the
finite-volume problem particularly in $0\nu2\beta$ decay $nn\to ppee$ as we will
explain below. A similar idea to use the massive photon as an infrared regularization
scheme for lattice QCQ+QED can be found in Ref.~\cite{Endres:2015gda}.

\subsection{\boldmath $2\nu2\beta$ decay $nn\to ppee\nu\nu$}

The pioneering lattice QCD calculation of $nn\to ppee\nu\nu$ has been performed by
NPLQCD collaboration~\cite{Shanahan:2017bgi,Tiburzi:2017iux}.
At the physical pion mass,
it is well known that the ${^1}S_0$ is a scattering state while ${^3}S_1$ is a shallow bound
state below the threshold and a scattering state above the threshold. In general,
we can treat the shallow bound state as a two-body system and use the 
finite-volume formula, Eq.~(\ref{eq:correction_formula}),
to relate the lattice results of finite-volume $nn\to ppee\nu\nu$
matrix element
to the infinite-volume decay amplitude.

\subsection{\boldmath $0\nu2\beta$ decay $nn\to ppee$}

The finite-volume problem for $0\nu2\beta$ decay $nn\to ppee$ is more
complicated for two reasons. First, the neutrino, proton and neutron in the
low-lying intermediate states form a three-body system. Second, the massless
neutrino enclosed in a finite-size box results in an additional power-law
finite-volume effect. Although Ref.~\cite{Feng:2018qpx} developed the infinite-volume
reconstruction method to eliminate the power-law finite-volume effects for 
the system with a massless photon and a stable hadron in the low-lying
intermediate state,
it is much harder to do this
for a system with a massless neutrino and two hadrons in the intermediate state.

Pointing out by Ref.~\cite{Cirigliano:2018hja},
a leading-order, short-range contribution needs to be introduced in the
$\chi$EFT study of the $nn\to
ppee$ decay. Such short-range contribution breaks down Weinberg's power-counting
scheme.
New local operators need to be introduced in the effective action to account for this contribution.
Our goal of the lattice calculation is to calculate the low energy constants for these new local operators.
Fortunately these low energy constants are irrelevant with the ultrasoft
region where neutrino's energy is much smaller than the pion mass. Besides, the
ultrasoft information from the $nn\to ppee$ decay is not very useful for
the heavy-nuclei $0\nu2\beta$ decay. In that case,
the ultrasoft neutrino can feel the complete nucleus instead of just the
nucleons. 
One would rely on the ab initio many-body theory to treat the nuclei properly.

We thus propose to introduce a nonzero mass for neutrino
to remove the ultrasoft contribution.
For simplicity, the neutrino mass can be chosen the same as the pion mass. Such
choice would unavoidably introduce the unphysical effects. However, as far as
the lattice QCD calculation and the $\chi$EFT use the same
unphysical neutrino mass,
the low energy constants can be determined in a clean way. Compared to the other
IR regulator such as the QED$_{\mathrm{L}}$ technique, introducing the massive
neutrino is relatively simpler for $\chi$EFT.
As far as the nonzero neutrino mass is introduced, at the threshold
of dibaryon, the three particles in the intermediate state cannot be on shell
simultaneously. Thus one can effectively treat the double beta decay as a 
$2\xrightarrow[]{J}2$ system with the current $J$ given by two weak operators.
The formula in Eq.~(\ref{eq:2to2}) can be applied to this case.

\section{Conclusion}

In this work we derive the finite-volume formula which connects a
$2\xrightarrow[]{H_I+H_I}2$ transition amplitudes in the finite volume to the
physical amplitudes in the infinite volume. We discuss the cases with
low-lying intermediate state consisting of two scattering hadrons or single stable
hadron. Using the idea originally proposed
by Lellouch and L\"uscher the derivation is simple and straightforward. As a
side product, we reproduce the finite-volume formalism for $2\xrightarrow[]{J}2$
transition previously obtained by Ref.~\cite{Briceno:2015tza}.

We discuss the application of the finite-volume formula of the
$2\xrightarrow[]{H_I+H_I}2$ transition to the lattice QCD calculation of the double
beta decay. In the case of $nn\to ppee$ decay, we propose to use the massive
neutrino to avoid the complication of the finite-volume problem induced by the
long-range massless neutrino.

\begin{acknowledgments}

We gratefully acknowledge many helpful discussions with our colleagues from the
RBC-UKQCD Collaboration. We warmly thank N.~H.~Christ, V. Cirigliano, W.~Dekens,
E.~Mereghetti and A. Walker-loud for useful discussion.
X.F., Z.Y.W. and Z.Z. were supported in part by NSFC of China under Grant No. 11775002.
L.C.J acknowledges support by DOE grant DE-SC0010339.
\end{acknowledgments}

\bibliography{paper}

\begin{thebibliography}{94}%
\makeatletter
\providecommand \@ifxundefined [1]{%
 \@ifx{#1\undefined}
}%
\providecommand \@ifnum [1]{%
 \ifnum #1\expandafter \@firstoftwo
 \else \expandafter \@secondoftwo
 \fi
}%
\providecommand \@ifx [1]{%
 \ifx #1\expandafter \@firstoftwo
 \else \expandafter \@secondoftwo
 \fi
}%
\providecommand \natexlab [1]{#1}%
\providecommand \enquote  [1]{``#1''}%
\providecommand \bibnamefont  [1]{#1}%
\providecommand \bibfnamefont [1]{#1}%
\providecommand \citenamefont [1]{#1}%
\providecommand \href@noop [0]{\@secondoftwo}%
\providecommand \href [0]{\begingroup \@sanitize@url \@href}%
\providecommand \@href[1]{\@@startlink{#1}\@@href}%
\providecommand \@@href[1]{\endgroup#1\@@endlink}%
\providecommand \@sanitize@url [0]{\catcode `\\12\catcode `\$12\catcode
  `\&12\catcode `\#12\catcode `\^12\catcode `\_12\catcode `\%12\relax}%
\providecommand \@@startlink[1]{}%
\providecommand \@@endlink[0]{}%
\providecommand \url  [0]{\begingroup\@sanitize@url \@url }%
\providecommand \@url [1]{\endgroup\@href {#1}{\urlprefix }}%
\providecommand \urlprefix  [0]{URL }%
\providecommand \Eprint [0]{\href }%
\providecommand \doibase [0]{http://dx.doi.org/}%
\providecommand \selectlanguage [0]{\@gobble}%
\providecommand \bibinfo  [0]{\@secondoftwo}%
\providecommand \bibfield  [0]{\@secondoftwo}%
\providecommand \translation [1]{[#1]}%
\providecommand \BibitemOpen [0]{}%
\providecommand \bibitemStop [0]{}%
\providecommand \bibitemNoStop [0]{.\EOS\space}%
\providecommand \EOS [0]{\spacefactor3000\relax}%
\providecommand \BibitemShut  [1]{\csname bibitem#1\endcsname}%
\let\auto@bib@innerbib\@empty
%</preamble>
\bibitem [{\citenamefont {Briceño}\ and\ \citenamefont
  {Hansen}(2016)}]{Briceno:2015tza}%
  \BibitemOpen
  \bibfield  {author} {\bibinfo {author} {\bibfnamefont {R.~A.}\ \bibnamefont
  {Briceño}}\ and\ \bibinfo {author} {\bibfnamefont {M.~T.}\ \bibnamefont
  {Hansen}},\ }\href {\doibase 10.1103/PhysRevD.94.013008} {\bibfield
  {journal} {\bibinfo  {journal} {Phys. Rev.}\ }\textbf {\bibinfo {volume}
  {D94}},\ \bibinfo {pages} {013008} (\bibinfo {year} {2016})},\ \Eprint
  {http://arxiv.org/abs/1509.08507} {arXiv:1509.08507 [hep-lat]} \BibitemShut
  {NoStop}%
%%CITATION = ARXIV:1509.08507;%%
\bibitem [{\citenamefont {Luscher}(1986)}]{Luscher:1986pf}%
  \BibitemOpen
  \bibfield  {author} {\bibinfo {author} {\bibfnamefont {M.}~\bibnamefont
  {Luscher}},\ }\href {\doibase 10.1007/BF01211097} {\bibfield  {journal}
  {\bibinfo  {journal} {Commun. Math. Phys.}\ }\textbf {\bibinfo {volume}
  {105}},\ \bibinfo {pages} {153} (\bibinfo {year} {1986})}\BibitemShut
  {NoStop}%
%%CITATION = CMPHA,105,153;%%
\bibitem [{\citenamefont {Luscher}(1991{\natexlab{a}})}]{Luscher:1990ux}%
  \BibitemOpen
  \bibfield  {author} {\bibinfo {author} {\bibfnamefont {M.}~\bibnamefont
  {Luscher}},\ }\href {\doibase 10.1016/0550-3213(91)90366-6} {\bibfield
  {journal} {\bibinfo  {journal} {Nucl. Phys.}\ }\textbf {\bibinfo {volume}
  {B354}},\ \bibinfo {pages} {531} (\bibinfo {year}
  {1991}{\natexlab{a}})}\BibitemShut {NoStop}%
%%CITATION = NUPHA,B354,531;%%
\bibitem [{\citenamefont {Luscher}(1991{\natexlab{b}})}]{Luscher:1991cf}%
  \BibitemOpen
  \bibfield  {author} {\bibinfo {author} {\bibfnamefont {M.}~\bibnamefont
  {Luscher}},\ }\href {\doibase 10.1016/0550-3213(91)90584-K} {\bibfield
  {journal} {\bibinfo  {journal} {Nucl. Phys.}\ }\textbf {\bibinfo {volume}
  {B364}},\ \bibinfo {pages} {237} (\bibinfo {year}
  {1991}{\natexlab{b}})}\BibitemShut {NoStop}%
%%CITATION = NUPHA,B364,237;%%
\bibitem [{\citenamefont {Yamazaki}\ \emph {et~al.}(2004)\citenamefont
  {Yamazaki} \emph {et~al.}}]{Yamazaki:2004qb}%
  \BibitemOpen
  \bibfield  {author} {\bibinfo {author} {\bibfnamefont {T.}~\bibnamefont
  {Yamazaki}} \emph {et~al.} (\bibinfo {collaboration} {CP-PACS}),\ }\href
  {\doibase 10.1103/PhysRevD.70.074513} {\bibfield  {journal} {\bibinfo
  {journal} {Phys. Rev.}\ }\textbf {\bibinfo {volume} {D70}},\ \bibinfo {pages}
  {074513} (\bibinfo {year} {2004})},\ \Eprint
  {http://arxiv.org/abs/hep-lat/0402025} {arXiv:hep-lat/0402025 [hep-lat]}
  \BibitemShut {NoStop}%
%%CITATION = HEP-LAT/0402025;%%
\bibitem [{\citenamefont {Beane}\ \emph {et~al.}(2006)\citenamefont {Beane},
  \citenamefont {Bedaque}, \citenamefont {Orginos},\ and\ \citenamefont
  {Savage}}]{Beane:2005rj}%
  \BibitemOpen
  \bibfield  {author} {\bibinfo {author} {\bibfnamefont {S.~R.}\ \bibnamefont
  {Beane}}, \bibinfo {author} {\bibfnamefont {P.~F.}\ \bibnamefont {Bedaque}},
  \bibinfo {author} {\bibfnamefont {K.}~\bibnamefont {Orginos}}, \ and\
  \bibinfo {author} {\bibfnamefont {M.~J.}\ \bibnamefont {Savage}} (\bibinfo
  {collaboration} {NPLQCD}),\ }\href {\doibase 10.1103/PhysRevD.73.054503}
  {\bibfield  {journal} {\bibinfo  {journal} {Phys. Rev.}\ }\textbf {\bibinfo
  {volume} {D73}},\ \bibinfo {pages} {054503} (\bibinfo {year} {2006})},\
  \Eprint {http://arxiv.org/abs/hep-lat/0506013} {arXiv:hep-lat/0506013
  [hep-lat]} \BibitemShut {NoStop}%
%%CITATION = HEP-LAT/0506013;%%
\bibitem [{\citenamefont {Beane}\ \emph {et~al.}(2008)\citenamefont {Beane},
  \citenamefont {Luu}, \citenamefont {Orginos}, \citenamefont {Parreno},
  \citenamefont {Savage}, \citenamefont {Torok},\ and\ \citenamefont
  {Walker-Loud}}]{Beane:2007xs}%
  \BibitemOpen
  \bibfield  {author} {\bibinfo {author} {\bibfnamefont {S.~R.}\ \bibnamefont
  {Beane}}, \bibinfo {author} {\bibfnamefont {T.~C.}\ \bibnamefont {Luu}},
  \bibinfo {author} {\bibfnamefont {K.}~\bibnamefont {Orginos}}, \bibinfo
  {author} {\bibfnamefont {A.}~\bibnamefont {Parreno}}, \bibinfo {author}
  {\bibfnamefont {M.~J.}\ \bibnamefont {Savage}}, \bibinfo {author}
  {\bibfnamefont {A.}~\bibnamefont {Torok}}, \ and\ \bibinfo {author}
  {\bibfnamefont {A.}~\bibnamefont {Walker-Loud}},\ }\href {\doibase
  10.1103/PhysRevD.77.014505} {\bibfield  {journal} {\bibinfo  {journal} {Phys.
  Rev.}\ }\textbf {\bibinfo {volume} {D77}},\ \bibinfo {pages} {014505}
  (\bibinfo {year} {2008})},\ \Eprint {http://arxiv.org/abs/0706.3026}
  {arXiv:0706.3026 [hep-lat]} \BibitemShut {NoStop}%
%%CITATION = ARXIV:0706.3026;%%
\bibitem [{\citenamefont {Feng}\ \emph {et~al.}(2010)\citenamefont {Feng},
  \citenamefont {Jansen},\ and\ \citenamefont {Renner}}]{Feng:2009ij}%
  \BibitemOpen
  \bibfield  {author} {\bibinfo {author} {\bibfnamefont {X.}~\bibnamefont
  {Feng}}, \bibinfo {author} {\bibfnamefont {K.}~\bibnamefont {Jansen}}, \ and\
  \bibinfo {author} {\bibfnamefont {D.~B.}\ \bibnamefont {Renner}},\ }\href
  {\doibase 10.1016/j.physletb.2010.01.018} {\bibfield  {journal} {\bibinfo
  {journal} {Phys. Lett. B}\ }\textbf {\bibinfo {volume} {684}},\ \bibinfo
  {pages} {268} (\bibinfo {year} {2010})},\ \Eprint
  {http://arxiv.org/abs/0909.3255} {arXiv:0909.3255 [hep-lat]} \BibitemShut
  {NoStop}%
\bibitem [{\citenamefont {Dudek}\ \emph {et~al.}(2011)\citenamefont {Dudek},
  \citenamefont {Edwards}, \citenamefont {Peardon}, \citenamefont {Richards},\
  and\ \citenamefont {Thomas}}]{Dudek:2010ew}%
  \BibitemOpen
  \bibfield  {author} {\bibinfo {author} {\bibfnamefont {J.~J.}\ \bibnamefont
  {Dudek}}, \bibinfo {author} {\bibfnamefont {R.~G.}\ \bibnamefont {Edwards}},
  \bibinfo {author} {\bibfnamefont {M.~J.}\ \bibnamefont {Peardon}}, \bibinfo
  {author} {\bibfnamefont {D.~G.}\ \bibnamefont {Richards}}, \ and\ \bibinfo
  {author} {\bibfnamefont {C.~E.}\ \bibnamefont {Thomas}},\ }\href {\doibase
  10.1103/PhysRevD.83.071504} {\bibfield  {journal} {\bibinfo  {journal} {Phys.
  Rev.}\ }\textbf {\bibinfo {volume} {D83}},\ \bibinfo {pages} {071504}
  (\bibinfo {year} {2011})},\ \Eprint {http://arxiv.org/abs/1011.6352}
  {arXiv:1011.6352 [hep-ph]} \BibitemShut {NoStop}%
%%CITATION = ARXIV:1011.6352;%%
\bibitem [{\citenamefont {Beane}\ \emph {et~al.}(2012)\citenamefont {Beane},
  \citenamefont {Chang}, \citenamefont {Detmold}, \citenamefont {Lin},
  \citenamefont {Luu}, \citenamefont {Orginos}, \citenamefont {Parreno},
  \citenamefont {Savage}, \citenamefont {Torok},\ and\ \citenamefont
  {Walker-Loud}}]{Beane:2011sc}%
  \BibitemOpen
  \bibfield  {author} {\bibinfo {author} {\bibfnamefont {S.~R.}\ \bibnamefont
  {Beane}}, \bibinfo {author} {\bibfnamefont {E.}~\bibnamefont {Chang}},
  \bibinfo {author} {\bibfnamefont {W.}~\bibnamefont {Detmold}}, \bibinfo
  {author} {\bibfnamefont {H.~W.}\ \bibnamefont {Lin}}, \bibinfo {author}
  {\bibfnamefont {T.~C.}\ \bibnamefont {Luu}}, \bibinfo {author} {\bibfnamefont
  {K.}~\bibnamefont {Orginos}}, \bibinfo {author} {\bibfnamefont
  {A.}~\bibnamefont {Parreno}}, \bibinfo {author} {\bibfnamefont {M.~J.}\
  \bibnamefont {Savage}}, \bibinfo {author} {\bibfnamefont {A.}~\bibnamefont
  {Torok}}, \ and\ \bibinfo {author} {\bibfnamefont {A.}~\bibnamefont
  {Walker-Loud}} (\bibinfo {collaboration} {NPLQCD}),\ }\href {\doibase
  10.1103/PhysRevD.85.034505} {\bibfield  {journal} {\bibinfo  {journal} {Phys.
  Rev.}\ }\textbf {\bibinfo {volume} {D85}},\ \bibinfo {pages} {034505}
  (\bibinfo {year} {2012})},\ \Eprint {http://arxiv.org/abs/1107.5023}
  {arXiv:1107.5023 [hep-lat]} \BibitemShut {NoStop}%
%%CITATION = ARXIV:1107.5023;%%
\bibitem [{\citenamefont {Fu}(2013)}]{Fu:2013ffa}%
  \BibitemOpen
  \bibfield  {author} {\bibinfo {author} {\bibfnamefont {Z.}~\bibnamefont
  {Fu}},\ }\href {\doibase 10.1103/PhysRevD.87.074501} {\bibfield  {journal}
  {\bibinfo  {journal} {Phys. Rev.}\ }\textbf {\bibinfo {volume} {D87}},\
  \bibinfo {pages} {074501} (\bibinfo {year} {2013})},\ \Eprint
  {http://arxiv.org/abs/1303.0517} {arXiv:1303.0517 [hep-lat]} \BibitemShut
  {NoStop}%
%%CITATION = ARXIV:1303.0517;%%
\bibitem [{\citenamefont {Sasaki}\ \emph {et~al.}(2014)\citenamefont {Sasaki},
  \citenamefont {Ishizuka}, \citenamefont {Oka},\ and\ \citenamefont
  {Yamazaki}}]{Sasaki:2013vxa}%
  \BibitemOpen
  \bibfield  {author} {\bibinfo {author} {\bibfnamefont {K.}~\bibnamefont
  {Sasaki}}, \bibinfo {author} {\bibfnamefont {N.}~\bibnamefont {Ishizuka}},
  \bibinfo {author} {\bibfnamefont {M.}~\bibnamefont {Oka}}, \ and\ \bibinfo
  {author} {\bibfnamefont {T.}~\bibnamefont {Yamazaki}} (\bibinfo
  {collaboration} {PACS-CS}),\ }\href {\doibase 10.1103/PhysRevD.89.054502}
  {\bibfield  {journal} {\bibinfo  {journal} {Phys. Rev.}\ }\textbf {\bibinfo
  {volume} {D89}},\ \bibinfo {pages} {054502} (\bibinfo {year} {2014})},\
  \Eprint {http://arxiv.org/abs/1311.7226} {arXiv:1311.7226 [hep-lat]}
  \BibitemShut {NoStop}%
%%CITATION = ARXIV:1311.7226;%%
\bibitem [{\citenamefont {Dudek}\ \emph {et~al.}(2012)\citenamefont {Dudek},
  \citenamefont {Edwards},\ and\ \citenamefont {Thomas}}]{Dudek:2012gj}%
  \BibitemOpen
  \bibfield  {author} {\bibinfo {author} {\bibfnamefont {J.~J.}\ \bibnamefont
  {Dudek}}, \bibinfo {author} {\bibfnamefont {R.~G.}\ \bibnamefont {Edwards}},
  \ and\ \bibinfo {author} {\bibfnamefont {C.~E.}\ \bibnamefont {Thomas}},\
  }\href {\doibase 10.1103/PhysRevD.86.034031} {\bibfield  {journal} {\bibinfo
  {journal} {Phys. Rev.}\ }\textbf {\bibinfo {volume} {D86}},\ \bibinfo {pages}
  {034031} (\bibinfo {year} {2012})},\ \Eprint {http://arxiv.org/abs/1203.6041}
  {arXiv:1203.6041 [hep-ph]} \BibitemShut {NoStop}%
%%CITATION = ARXIV:1203.6041;%%
\bibitem [{\citenamefont {Helmes}\ \emph {et~al.}(2015)\citenamefont {Helmes},
  \citenamefont {Jost}, \citenamefont {Knippschild}, \citenamefont {Liu},
  \citenamefont {Liu}, \citenamefont {Liu}, \citenamefont {Urbach},
  \citenamefont {Ueding}, \citenamefont {Wang},\ and\ \citenamefont
  {Werner}}]{Helmes:2015gla}%
  \BibitemOpen
  \bibfield  {author} {\bibinfo {author} {\bibfnamefont {C.}~\bibnamefont
  {Helmes}}, \bibinfo {author} {\bibfnamefont {C.}~\bibnamefont {Jost}},
  \bibinfo {author} {\bibfnamefont {B.}~\bibnamefont {Knippschild}}, \bibinfo
  {author} {\bibfnamefont {C.}~\bibnamefont {Liu}}, \bibinfo {author}
  {\bibfnamefont {J.}~\bibnamefont {Liu}}, \bibinfo {author} {\bibfnamefont
  {L.}~\bibnamefont {Liu}}, \bibinfo {author} {\bibfnamefont {C.}~\bibnamefont
  {Urbach}}, \bibinfo {author} {\bibfnamefont {M.}~\bibnamefont {Ueding}},
  \bibinfo {author} {\bibfnamefont {Z.}~\bibnamefont {Wang}}, \ and\ \bibinfo
  {author} {\bibfnamefont {M.}~\bibnamefont {Werner}} (\bibinfo {collaboration}
  {ETM}),\ }\href {\doibase 10.1007/JHEP09(2015)109} {\bibfield  {journal}
  {\bibinfo  {journal} {JHEP}\ }\textbf {\bibinfo {volume} {09}},\ \bibinfo
  {pages} {109} (\bibinfo {year} {2015})},\ \Eprint
  {http://arxiv.org/abs/1506.00408} {arXiv:1506.00408 [hep-lat]} \BibitemShut
  {NoStop}%
%%CITATION = ARXIV:1506.00408;%%
\bibitem [{\citenamefont {Culver}\ \emph {et~al.}(2019)\citenamefont {Culver},
  \citenamefont {Mai}, \citenamefont {Alexandru}, \citenamefont {Döring},\
  and\ \citenamefont {Lee}}]{Culver:2019qtx}%
  \BibitemOpen
  \bibfield  {author} {\bibinfo {author} {\bibfnamefont {C.}~\bibnamefont
  {Culver}}, \bibinfo {author} {\bibfnamefont {M.}~\bibnamefont {Mai}},
  \bibinfo {author} {\bibfnamefont {A.}~\bibnamefont {Alexandru}}, \bibinfo
  {author} {\bibfnamefont {M.}~\bibnamefont {Döring}}, \ and\ \bibinfo
  {author} {\bibfnamefont {F.~X.}\ \bibnamefont {Lee}},\ }\href {\doibase
  10.1103/PhysRevD.100.034509} {\bibfield  {journal} {\bibinfo  {journal}
  {Phys. Rev.}\ }\textbf {\bibinfo {volume} {D100}},\ \bibinfo {pages} {034509}
  (\bibinfo {year} {2019})},\ \Eprint {http://arxiv.org/abs/1905.10202}
  {arXiv:1905.10202 [hep-lat]} \BibitemShut {NoStop}%
%%CITATION = ARXIV:1905.10202;%%
\bibitem [{\citenamefont {Aoki}\ \emph {et~al.}(2007)\citenamefont {Aoki} \emph
  {et~al.}}]{Aoki:2007rd}%
  \BibitemOpen
  \bibfield  {author} {\bibinfo {author} {\bibfnamefont {S.}~\bibnamefont
  {Aoki}} \emph {et~al.} (\bibinfo {collaboration} {CP-PACS}),\ }\href
  {\doibase 10.1103/PhysRevD.76.094506} {\bibfield  {journal} {\bibinfo
  {journal} {Phys. Rev.}\ }\textbf {\bibinfo {volume} {D76}},\ \bibinfo {pages}
  {094506} (\bibinfo {year} {2007})},\ \Eprint {http://arxiv.org/abs/0708.3705}
  {arXiv:0708.3705 [hep-lat]} \BibitemShut {NoStop}%
%%CITATION = ARXIV:0708.3705;%%
\bibitem [{\citenamefont {Feng}\ \emph {et~al.}(2011)\citenamefont {Feng},
  \citenamefont {Jansen},\ and\ \citenamefont {Renner}}]{Feng:2010es}%
  \BibitemOpen
  \bibfield  {author} {\bibinfo {author} {\bibfnamefont {X.}~\bibnamefont
  {Feng}}, \bibinfo {author} {\bibfnamefont {K.}~\bibnamefont {Jansen}}, \ and\
  \bibinfo {author} {\bibfnamefont {D.~B.}\ \bibnamefont {Renner}},\ }\href
  {\doibase 10.1103/PhysRevD.83.094505} {\bibfield  {journal} {\bibinfo
  {journal} {Phys. Rev.}\ }\textbf {\bibinfo {volume} {D83}},\ \bibinfo {pages}
  {094505} (\bibinfo {year} {2011})},\ \Eprint {http://arxiv.org/abs/1011.5288}
  {arXiv:1011.5288 [hep-lat]} \BibitemShut {NoStop}%
%%CITATION = ARXIV:1011.5288;%%
\bibitem [{\citenamefont {Lang}\ \emph {et~al.}(2011)\citenamefont {Lang},
  \citenamefont {Mohler}, \citenamefont {Prelovsek},\ and\ \citenamefont
  {Vidmar}}]{Lang:2011mn}%
  \BibitemOpen
  \bibfield  {author} {\bibinfo {author} {\bibfnamefont {C.~B.}\ \bibnamefont
  {Lang}}, \bibinfo {author} {\bibfnamefont {D.}~\bibnamefont {Mohler}},
  \bibinfo {author} {\bibfnamefont {S.}~\bibnamefont {Prelovsek}}, \ and\
  \bibinfo {author} {\bibfnamefont {M.}~\bibnamefont {Vidmar}},\ }\href
  {\doibase 10.1103/PhysRevD.89.059903, 10.1103/PhysRevD.84.054503} {\bibfield
  {journal} {\bibinfo  {journal} {Phys. Rev.}\ }\textbf {\bibinfo {volume}
  {D84}},\ \bibinfo {pages} {054503} (\bibinfo {year} {2011})},\ \bibinfo
  {note} {[Erratum: Phys. Rev.D89,no.5,059903(2014)]},\ \Eprint
  {http://arxiv.org/abs/1105.5636} {arXiv:1105.5636 [hep-lat]} \BibitemShut
  {NoStop}%
%%CITATION = ARXIV:1105.5636;%%
\bibitem [{\citenamefont {Aoki}\ \emph
  {et~al.}(2011{\natexlab{a}})\citenamefont {Aoki} \emph
  {et~al.}}]{Aoki:2011yj}%
  \BibitemOpen
  \bibfield  {author} {\bibinfo {author} {\bibfnamefont {S.}~\bibnamefont
  {Aoki}} \emph {et~al.} (\bibinfo {collaboration} {CS}),\ }\href {\doibase
  10.1103/PhysRevD.84.094505} {\bibfield  {journal} {\bibinfo  {journal} {Phys.
  Rev.}\ }\textbf {\bibinfo {volume} {D84}},\ \bibinfo {pages} {094505}
  (\bibinfo {year} {2011}{\natexlab{a}})},\ \Eprint
  {http://arxiv.org/abs/1106.5365} {arXiv:1106.5365 [hep-lat]} \BibitemShut
  {NoStop}%
%%CITATION = ARXIV:1106.5365;%%
\bibitem [{\citenamefont {Pelissier}\ and\ \citenamefont
  {Alexandru}(2013)}]{Pelissier:2012pi}%
  \BibitemOpen
  \bibfield  {author} {\bibinfo {author} {\bibfnamefont {C.}~\bibnamefont
  {Pelissier}}\ and\ \bibinfo {author} {\bibfnamefont {A.}~\bibnamefont
  {Alexandru}},\ }\href {\doibase 10.1103/PhysRevD.87.014503} {\bibfield
  {journal} {\bibinfo  {journal} {Phys. Rev.}\ }\textbf {\bibinfo {volume}
  {D87}},\ \bibinfo {pages} {014503} (\bibinfo {year} {2013})},\ \Eprint
  {http://arxiv.org/abs/1211.0092} {arXiv:1211.0092 [hep-lat]} \BibitemShut
  {NoStop}%
%%CITATION = ARXIV:1211.0092;%%
\bibitem [{\citenamefont {Dudek}\ \emph {et~al.}(2013)\citenamefont {Dudek},
  \citenamefont {Edwards},\ and\ \citenamefont {Thomas}}]{Dudek:2012xn}%
  \BibitemOpen
  \bibfield  {author} {\bibinfo {author} {\bibfnamefont {J.~J.}\ \bibnamefont
  {Dudek}}, \bibinfo {author} {\bibfnamefont {R.~G.}\ \bibnamefont {Edwards}},
  \ and\ \bibinfo {author} {\bibfnamefont {C.~E.}\ \bibnamefont {Thomas}}
  (\bibinfo {collaboration} {Hadron Spectrum}),\ }\href {\doibase
  10.1103/PhysRevD.87.034505, 10.1103/PhysRevD.90.099902} {\bibfield  {journal}
  {\bibinfo  {journal} {Phys. Rev.}\ }\textbf {\bibinfo {volume} {D87}},\
  \bibinfo {pages} {034505} (\bibinfo {year} {2013})},\ \bibinfo {note}
  {[Erratum: Phys. Rev.D90,no.9,099902(2014)]},\ \Eprint
  {http://arxiv.org/abs/1212.0830} {arXiv:1212.0830 [hep-ph]} \BibitemShut
  {NoStop}%
%%CITATION = ARXIV:1212.0830;%%
\bibitem [{\citenamefont {Feng}\ \emph {et~al.}(2015)\citenamefont {Feng},
  \citenamefont {Aoki}, \citenamefont {Hashimoto},\ and\ \citenamefont
  {Kaneko}}]{Feng:2014gba}%
  \BibitemOpen
  \bibfield  {author} {\bibinfo {author} {\bibfnamefont {X.}~\bibnamefont
  {Feng}}, \bibinfo {author} {\bibfnamefont {S.}~\bibnamefont {Aoki}}, \bibinfo
  {author} {\bibfnamefont {S.}~\bibnamefont {Hashimoto}}, \ and\ \bibinfo
  {author} {\bibfnamefont {T.}~\bibnamefont {Kaneko}},\ }\href {\doibase
  10.1103/PhysRevD.91.054504} {\bibfield  {journal} {\bibinfo  {journal} {Phys.
  Rev.}\ }\textbf {\bibinfo {volume} {D91}},\ \bibinfo {pages} {054504}
  (\bibinfo {year} {2015})},\ \Eprint {http://arxiv.org/abs/1412.6319}
  {arXiv:1412.6319 [hep-lat]} \BibitemShut {NoStop}%
%%CITATION = ARXIV:1412.6319;%%
\bibitem [{\citenamefont {Wilson}\ \emph {et~al.}(2015)\citenamefont {Wilson},
  \citenamefont {Briceno}, \citenamefont {Dudek}, \citenamefont {Edwards},\
  and\ \citenamefont {Thomas}}]{Wilson:2015dqa}%
  \BibitemOpen
  \bibfield  {author} {\bibinfo {author} {\bibfnamefont {D.~J.}\ \bibnamefont
  {Wilson}}, \bibinfo {author} {\bibfnamefont {R.~A.}\ \bibnamefont {Briceno}},
  \bibinfo {author} {\bibfnamefont {J.~J.}\ \bibnamefont {Dudek}}, \bibinfo
  {author} {\bibfnamefont {R.~G.}\ \bibnamefont {Edwards}}, \ and\ \bibinfo
  {author} {\bibfnamefont {C.~E.}\ \bibnamefont {Thomas}},\ }\href {\doibase
  10.1103/PhysRevD.92.094502} {\bibfield  {journal} {\bibinfo  {journal} {Phys.
  Rev.}\ }\textbf {\bibinfo {volume} {D92}},\ \bibinfo {pages} {094502}
  (\bibinfo {year} {2015})},\ \Eprint {http://arxiv.org/abs/1507.02599}
  {arXiv:1507.02599 [hep-ph]} \BibitemShut {NoStop}%
%%CITATION = ARXIV:1507.02599;%%
\bibitem [{\citenamefont {Bali}\ \emph {et~al.}(2016)\citenamefont {Bali},
  \citenamefont {Collins}, \citenamefont {Cox}, \citenamefont {Donald},
  \citenamefont {Göckeler}, \citenamefont {Lang},\ and\ \citenamefont
  {Schäfer}}]{Bali:2015gji}%
  \BibitemOpen
  \bibfield  {author} {\bibinfo {author} {\bibfnamefont {G.~S.}\ \bibnamefont
  {Bali}}, \bibinfo {author} {\bibfnamefont {S.}~\bibnamefont {Collins}},
  \bibinfo {author} {\bibfnamefont {A.}~\bibnamefont {Cox}}, \bibinfo {author}
  {\bibfnamefont {G.}~\bibnamefont {Donald}}, \bibinfo {author} {\bibfnamefont
  {M.}~\bibnamefont {Göckeler}}, \bibinfo {author} {\bibfnamefont {C.~B.}\
  \bibnamefont {Lang}}, \ and\ \bibinfo {author} {\bibfnamefont
  {A.}~\bibnamefont {Schäfer}} (\bibinfo {collaboration} {RQCD}),\ }\href
  {\doibase 10.1103/PhysRevD.93.054509} {\bibfield  {journal} {\bibinfo
  {journal} {Phys. Rev.}\ }\textbf {\bibinfo {volume} {D93}},\ \bibinfo {pages}
  {054509} (\bibinfo {year} {2016})},\ \Eprint
  {http://arxiv.org/abs/1512.08678} {arXiv:1512.08678 [hep-lat]} \BibitemShut
  {NoStop}%
%%CITATION = ARXIV:1512.08678;%%
\bibitem [{\citenamefont {Bulava}\ \emph {et~al.}(2016)\citenamefont {Bulava},
  \citenamefont {Fahy}, \citenamefont {Hörz}, \citenamefont {Juge},
  \citenamefont {Morningstar},\ and\ \citenamefont {Wong}}]{Bulava:2016mks}%
  \BibitemOpen
  \bibfield  {author} {\bibinfo {author} {\bibfnamefont {J.}~\bibnamefont
  {Bulava}}, \bibinfo {author} {\bibfnamefont {B.}~\bibnamefont {Fahy}},
  \bibinfo {author} {\bibfnamefont {B.}~\bibnamefont {Hörz}}, \bibinfo
  {author} {\bibfnamefont {K.~J.}\ \bibnamefont {Juge}}, \bibinfo {author}
  {\bibfnamefont {C.}~\bibnamefont {Morningstar}}, \ and\ \bibinfo {author}
  {\bibfnamefont {C.~H.}\ \bibnamefont {Wong}},\ }\href {\doibase
  10.1016/j.nuclphysb.2016.07.024} {\bibfield  {journal} {\bibinfo  {journal}
  {Nucl. Phys.}\ }\textbf {\bibinfo {volume} {B910}},\ \bibinfo {pages} {842}
  (\bibinfo {year} {2016})},\ \Eprint {http://arxiv.org/abs/1604.05593}
  {arXiv:1604.05593 [hep-lat]} \BibitemShut {NoStop}%
%%CITATION = ARXIV:1604.05593;%%
\bibitem [{\citenamefont {Guo}\ \emph {et~al.}(2016)\citenamefont {Guo},
  \citenamefont {Alexandru}, \citenamefont {Molina},\ and\ \citenamefont
  {Döring}}]{Guo:2016zos}%
  \BibitemOpen
  \bibfield  {author} {\bibinfo {author} {\bibfnamefont {D.}~\bibnamefont
  {Guo}}, \bibinfo {author} {\bibfnamefont {A.}~\bibnamefont {Alexandru}},
  \bibinfo {author} {\bibfnamefont {R.}~\bibnamefont {Molina}}, \ and\ \bibinfo
  {author} {\bibfnamefont {M.}~\bibnamefont {Döring}},\ }\href {\doibase
  10.1103/PhysRevD.94.034501} {\bibfield  {journal} {\bibinfo  {journal} {Phys.
  Rev.}\ }\textbf {\bibinfo {volume} {D94}},\ \bibinfo {pages} {034501}
  (\bibinfo {year} {2016})},\ \Eprint {http://arxiv.org/abs/1605.03993}
  {arXiv:1605.03993 [hep-lat]} \BibitemShut {NoStop}%
%%CITATION = ARXIV:1605.03993;%%
\bibitem [{\citenamefont {Fu}\ and\ \citenamefont {Wang}(2016)}]{Fu:2016itp}%
  \BibitemOpen
  \bibfield  {author} {\bibinfo {author} {\bibfnamefont {Z.}~\bibnamefont
  {Fu}}\ and\ \bibinfo {author} {\bibfnamefont {L.}~\bibnamefont {Wang}},\
  }\href {\doibase 10.1103/PhysRevD.94.034505} {\bibfield  {journal} {\bibinfo
  {journal} {Phys. Rev.}\ }\textbf {\bibinfo {volume} {D94}},\ \bibinfo {pages}
  {034505} (\bibinfo {year} {2016})},\ \Eprint
  {http://arxiv.org/abs/1608.07478} {arXiv:1608.07478 [hep-lat]} \BibitemShut
  {NoStop}%
%%CITATION = ARXIV:1608.07478;%%
\bibitem [{\citenamefont {Alexandrou}\ \emph {et~al.}(2017)\citenamefont
  {Alexandrou}, \citenamefont {Leskovec}, \citenamefont {Meinel}, \citenamefont
  {Negele}, \citenamefont {Paul}, \citenamefont {Petschlies}, \citenamefont
  {Pochinsky}, \citenamefont {Rendon},\ and\ \citenamefont
  {Syritsyn}}]{Alexandrou:2017mpi}%
  \BibitemOpen
  \bibfield  {author} {\bibinfo {author} {\bibfnamefont {C.}~\bibnamefont
  {Alexandrou}}, \bibinfo {author} {\bibfnamefont {L.}~\bibnamefont
  {Leskovec}}, \bibinfo {author} {\bibfnamefont {S.}~\bibnamefont {Meinel}},
  \bibinfo {author} {\bibfnamefont {J.}~\bibnamefont {Negele}}, \bibinfo
  {author} {\bibfnamefont {S.}~\bibnamefont {Paul}}, \bibinfo {author}
  {\bibfnamefont {M.}~\bibnamefont {Petschlies}}, \bibinfo {author}
  {\bibfnamefont {A.}~\bibnamefont {Pochinsky}}, \bibinfo {author}
  {\bibfnamefont {G.}~\bibnamefont {Rendon}}, \ and\ \bibinfo {author}
  {\bibfnamefont {S.}~\bibnamefont {Syritsyn}},\ }\href {\doibase
  10.1103/PhysRevD.96.034525} {\bibfield  {journal} {\bibinfo  {journal} {Phys.
  Rev.}\ }\textbf {\bibinfo {volume} {D96}},\ \bibinfo {pages} {034525}
  (\bibinfo {year} {2017})},\ \Eprint {http://arxiv.org/abs/1704.05439}
  {arXiv:1704.05439 [hep-lat]} \BibitemShut {NoStop}%
%%CITATION = ARXIV:1704.05439;%%
\bibitem [{\citenamefont {Andersen}\ \emph {et~al.}(2019)\citenamefont
  {Andersen}, \citenamefont {Bulava}, \citenamefont {Hörz},\ and\
  \citenamefont {Morningstar}}]{Andersen:2018mau}%
  \BibitemOpen
  \bibfield  {author} {\bibinfo {author} {\bibfnamefont {C.}~\bibnamefont
  {Andersen}}, \bibinfo {author} {\bibfnamefont {J.}~\bibnamefont {Bulava}},
  \bibinfo {author} {\bibfnamefont {B.}~\bibnamefont {Hörz}}, \ and\ \bibinfo
  {author} {\bibfnamefont {C.}~\bibnamefont {Morningstar}},\ }\href {\doibase
  10.1016/j.nuclphysb.2018.12.018} {\bibfield  {journal} {\bibinfo  {journal}
  {Nucl. Phys.}\ }\textbf {\bibinfo {volume} {B939}},\ \bibinfo {pages} {145}
  (\bibinfo {year} {2019})},\ \Eprint {http://arxiv.org/abs/1808.05007}
  {arXiv:1808.05007 [hep-lat]} \BibitemShut {NoStop}%
%%CITATION = ARXIV:1808.05007;%%
\bibitem [{\citenamefont {Werner}\ \emph {et~al.}(2020)\citenamefont {Werner}
  \emph {et~al.}}]{Werner:2019hxc}%
  \BibitemOpen
  \bibfield  {author} {\bibinfo {author} {\bibfnamefont {M.}~\bibnamefont
  {Werner}} \emph {et~al.},\ }\href {\doibase 10.1140/epja/s10050-020-00057-4}
  {\bibfield  {journal} {\bibinfo  {journal} {Eur. Phys. J.}\ }\textbf
  {\bibinfo {volume} {A56}},\ \bibinfo {pages} {61} (\bibinfo {year} {2020})},\
  \Eprint {http://arxiv.org/abs/1907.01237} {arXiv:1907.01237 [hep-lat]}
  \BibitemShut {NoStop}%
%%CITATION = ARXIV:1907.01237;%%
\bibitem [{\citenamefont {Erben}\ \emph {et~al.}(2020)\citenamefont {Erben},
  \citenamefont {Green}, \citenamefont {Mohler},\ and\ \citenamefont
  {Wittig}}]{Erben:2019nmx}%
  \BibitemOpen
  \bibfield  {author} {\bibinfo {author} {\bibfnamefont {F.}~\bibnamefont
  {Erben}}, \bibinfo {author} {\bibfnamefont {J.~R.}\ \bibnamefont {Green}},
  \bibinfo {author} {\bibfnamefont {D.}~\bibnamefont {Mohler}}, \ and\ \bibinfo
  {author} {\bibfnamefont {H.}~\bibnamefont {Wittig}},\ }\href {\doibase
  10.1103/PhysRevD.101.054504} {\bibfield  {journal} {\bibinfo  {journal}
  {Phys. Rev.}\ }\textbf {\bibinfo {volume} {D101}},\ \bibinfo {pages} {054504}
  (\bibinfo {year} {2020})},\ \Eprint {http://arxiv.org/abs/1910.01083}
  {arXiv:1910.01083 [hep-lat]} \BibitemShut {NoStop}%
%%CITATION = ARXIV:1910.01083;%%
\bibitem [{\citenamefont {Liu}(2009)}]{Liu:2009uw}%
  \BibitemOpen
  \bibfield  {author} {\bibinfo {author} {\bibfnamefont {Q.}~\bibnamefont
  {Liu}} (\bibinfo {collaboration} {RBC, UKQCD}),\ }\bibfield  {booktitle}
  {\emph {\bibinfo {booktitle} {{Proceedings, 27th International Symposium on
  Lattice field theory (Lattice 2009): Beijing, P.R. China, July 26-31,
  2009}}},\ }\href {\doibase 10.22323/1.091.0101} {\bibfield  {journal}
  {\bibinfo  {journal} {PoS}\ }\textbf {\bibinfo {volume} {LAT2009}},\ \bibinfo
  {pages} {101} (\bibinfo {year} {2009})},\ \Eprint
  {http://arxiv.org/abs/0910.2658} {arXiv:0910.2658 [hep-lat]} \BibitemShut
  {NoStop}%
%%CITATION = ARXIV:0910.2658;%%
\bibitem [{\citenamefont {Briceno}\ \emph {et~al.}(2017)\citenamefont
  {Briceno}, \citenamefont {Dudek}, \citenamefont {Edwards},\ and\
  \citenamefont {Wilson}}]{Briceno:2016mjc}%
  \BibitemOpen
  \bibfield  {author} {\bibinfo {author} {\bibfnamefont {R.~A.}\ \bibnamefont
  {Briceno}}, \bibinfo {author} {\bibfnamefont {J.~J.}\ \bibnamefont {Dudek}},
  \bibinfo {author} {\bibfnamefont {R.~G.}\ \bibnamefont {Edwards}}, \ and\
  \bibinfo {author} {\bibfnamefont {D.~J.}\ \bibnamefont {Wilson}},\ }\href
  {\doibase 10.1103/PhysRevLett.118.022002} {\bibfield  {journal} {\bibinfo
  {journal} {Phys. Rev. Lett.}\ }\textbf {\bibinfo {volume} {118}},\ \bibinfo
  {pages} {022002} (\bibinfo {year} {2017})},\ \Eprint
  {http://arxiv.org/abs/1607.05900} {arXiv:1607.05900 [hep-ph]} \BibitemShut
  {NoStop}%
%%CITATION = ARXIV:1607.05900;%%
\bibitem [{\citenamefont {Liu}\ \emph {et~al.}(2017)\citenamefont {Liu} \emph
  {et~al.}}]{Liu:2016cba}%
  \BibitemOpen
  \bibfield  {author} {\bibinfo {author} {\bibfnamefont {L.}~\bibnamefont
  {Liu}} \emph {et~al.},\ }\href {\doibase 10.1103/PhysRevD.96.054516}
  {\bibfield  {journal} {\bibinfo  {journal} {Phys. Rev.}\ }\textbf {\bibinfo
  {volume} {D96}},\ \bibinfo {pages} {054516} (\bibinfo {year} {2017})},\
  \Eprint {http://arxiv.org/abs/1612.02061} {arXiv:1612.02061 [hep-lat]}
  \BibitemShut {NoStop}%
%%CITATION = ARXIV:1612.02061;%%
\bibitem [{\citenamefont {Briceno}\ \emph
  {et~al.}(2018{\natexlab{a}})\citenamefont {Briceno}, \citenamefont {Dudek},
  \citenamefont {Edwards},\ and\ \citenamefont {Wilson}}]{Briceno:2017qmb}%
  \BibitemOpen
  \bibfield  {author} {\bibinfo {author} {\bibfnamefont {R.~A.}\ \bibnamefont
  {Briceno}}, \bibinfo {author} {\bibfnamefont {J.~J.}\ \bibnamefont {Dudek}},
  \bibinfo {author} {\bibfnamefont {R.~G.}\ \bibnamefont {Edwards}}, \ and\
  \bibinfo {author} {\bibfnamefont {D.~J.}\ \bibnamefont {Wilson}},\ }\href
  {\doibase 10.1103/PhysRevD.97.054513} {\bibfield  {journal} {\bibinfo
  {journal} {Phys. Rev.}\ }\textbf {\bibinfo {volume} {D97}},\ \bibinfo {pages}
  {054513} (\bibinfo {year} {2018}{\natexlab{a}})},\ \Eprint
  {http://arxiv.org/abs/1708.06667} {arXiv:1708.06667 [hep-lat]} \BibitemShut
  {NoStop}%
%%CITATION = ARXIV:1708.06667;%%
\bibitem [{\citenamefont {Fu}\ and\ \citenamefont {Chen}(2018)}]{Fu:2017apw}%
  \BibitemOpen
  \bibfield  {author} {\bibinfo {author} {\bibfnamefont {Z.}~\bibnamefont
  {Fu}}\ and\ \bibinfo {author} {\bibfnamefont {X.}~\bibnamefont {Chen}},\
  }\href {\doibase 10.1103/PhysRevD.98.014514} {\bibfield  {journal} {\bibinfo
  {journal} {Phys. Rev.}\ }\textbf {\bibinfo {volume} {D98}},\ \bibinfo {pages}
  {014514} (\bibinfo {year} {2018})},\ \Eprint
  {http://arxiv.org/abs/1712.02219} {arXiv:1712.02219 [hep-lat]} \BibitemShut
  {NoStop}%
%%CITATION = ARXIV:1712.02219;%%
\bibitem [{\citenamefont {Guo}\ \emph {et~al.}(2018)\citenamefont {Guo},
  \citenamefont {Alexandru}, \citenamefont {Molina}, \citenamefont {Mai},\ and\
  \citenamefont {Döring}}]{Guo:2018zss}%
  \BibitemOpen
  \bibfield  {author} {\bibinfo {author} {\bibfnamefont {D.}~\bibnamefont
  {Guo}}, \bibinfo {author} {\bibfnamefont {A.}~\bibnamefont {Alexandru}},
  \bibinfo {author} {\bibfnamefont {R.}~\bibnamefont {Molina}}, \bibinfo
  {author} {\bibfnamefont {M.}~\bibnamefont {Mai}}, \ and\ \bibinfo {author}
  {\bibfnamefont {M.}~\bibnamefont {Döring}},\ }\href {\doibase
  10.1103/PhysRevD.98.014507} {\bibfield  {journal} {\bibinfo  {journal} {Phys.
  Rev.}\ }\textbf {\bibinfo {volume} {D98}},\ \bibinfo {pages} {014507}
  (\bibinfo {year} {2018})},\ \Eprint {http://arxiv.org/abs/1803.02897}
  {arXiv:1803.02897 [hep-lat]} \BibitemShut {NoStop}%
%%CITATION = ARXIV:1803.02897;%%
\bibitem [{\citenamefont {Wang}\ and\ \citenamefont
  {Kelly}(2019)}]{Wang:2019nes}%
  \BibitemOpen
  \bibfield  {author} {\bibinfo {author} {\bibfnamefont {T.}~\bibnamefont
  {Wang}}\ and\ \bibinfo {author} {\bibfnamefont {C.}~\bibnamefont {Kelly}},\
  }\bibfield  {booktitle} {\emph {\bibinfo {booktitle} {{Proceedings, 36th
  International Symposium on Lattice Field Theory (Lattice 2018): East Lansing,
  MI, United States, July 22-28, 2018}}},\ }\href {\doibase
  10.22323/1.334.0276} {\bibfield  {journal} {\bibinfo  {journal} {PoS}\
  }\textbf {\bibinfo {volume} {LATTICE2018}},\ \bibinfo {pages} {276} (\bibinfo
  {year} {2019})}\BibitemShut {NoStop}%
%%CITATION = POSCI,LATTICE2018,276;%%
\bibitem [{\citenamefont {Briceno}\ \emph
  {et~al.}(2018{\natexlab{b}})\citenamefont {Briceno}, \citenamefont {Dudek},\
  and\ \citenamefont {Young}}]{Briceno:2017max}%
  \BibitemOpen
  \bibfield  {author} {\bibinfo {author} {\bibfnamefont {R.~A.}\ \bibnamefont
  {Briceno}}, \bibinfo {author} {\bibfnamefont {J.~J.}\ \bibnamefont {Dudek}},
  \ and\ \bibinfo {author} {\bibfnamefont {R.~D.}\ \bibnamefont {Young}},\
  }\href {\doibase 10.1103/RevModPhys.90.025001} {\bibfield  {journal}
  {\bibinfo  {journal} {Rev. Mod. Phys.}\ }\textbf {\bibinfo {volume} {90}},\
  \bibinfo {pages} {025001} (\bibinfo {year} {2018}{\natexlab{b}})},\ \Eprint
  {http://arxiv.org/abs/1706.06223} {arXiv:1706.06223 [hep-lat]} \BibitemShut
  {NoStop}%
\bibitem [{\citenamefont {Lellouch}\ and\ \citenamefont
  {Luscher}(2001)}]{Lellouch:2000pv}%
  \BibitemOpen
  \bibfield  {author} {\bibinfo {author} {\bibfnamefont {L.}~\bibnamefont
  {Lellouch}}\ and\ \bibinfo {author} {\bibfnamefont {M.}~\bibnamefont
  {Luscher}},\ }\href {\doibase 10.1007/s002200100410} {\bibfield  {journal}
  {\bibinfo  {journal} {Commun. Math. Phys.}\ }\textbf {\bibinfo {volume}
  {219}},\ \bibinfo {pages} {31} (\bibinfo {year} {2001})},\ \Eprint
  {http://arxiv.org/abs/hep-lat/0003023} {arXiv:hep-lat/0003023 [hep-lat]}
  \BibitemShut {NoStop}%
%%CITATION = HEP-LAT/0003023;%%
\bibitem [{\citenamefont {Meyer}(2011)}]{Meyer:2011um}%
  \BibitemOpen
  \bibfield  {author} {\bibinfo {author} {\bibfnamefont {H.~B.}\ \bibnamefont
  {Meyer}},\ }\href {\doibase 10.1103/PhysRevLett.107.072002} {\bibfield
  {journal} {\bibinfo  {journal} {Phys. Rev. Lett.}\ }\textbf {\bibinfo
  {volume} {107}},\ \bibinfo {pages} {072002} (\bibinfo {year} {2011})},\
  \Eprint {http://arxiv.org/abs/1105.1892} {arXiv:1105.1892 [hep-lat]}
  \BibitemShut {NoStop}%
\bibitem [{\citenamefont {Blum}\ \emph
  {et~al.}(2012{\natexlab{a}})\citenamefont {Blum} \emph
  {et~al.}}]{Blum:2011ng}%
  \BibitemOpen
  \bibfield  {author} {\bibinfo {author} {\bibfnamefont {T.}~\bibnamefont
  {Blum}} \emph {et~al.},\ }\href {\doibase 10.1103/PhysRevLett.108.141601}
  {\bibfield  {journal} {\bibinfo  {journal} {Phys. Rev. Lett.}\ }\textbf
  {\bibinfo {volume} {108}},\ \bibinfo {pages} {141601} (\bibinfo {year}
  {2012}{\natexlab{a}})},\ \Eprint {http://arxiv.org/abs/1111.1699}
  {arXiv:1111.1699 [hep-lat]} \BibitemShut {NoStop}%
%%CITATION = ARXIV:1111.1699;%%
\bibitem [{\citenamefont {Blum}\ \emph
  {et~al.}(2012{\natexlab{b}})\citenamefont {Blum} \emph
  {et~al.}}]{Blum:2012uk}%
  \BibitemOpen
  \bibfield  {author} {\bibinfo {author} {\bibfnamefont {T.}~\bibnamefont
  {Blum}} \emph {et~al.},\ }\href {\doibase 10.1103/PhysRevD.86.074513}
  {\bibfield  {journal} {\bibinfo  {journal} {Phys. Rev.}\ }\textbf {\bibinfo
  {volume} {D86}},\ \bibinfo {pages} {074513} (\bibinfo {year}
  {2012}{\natexlab{b}})},\ \Eprint {http://arxiv.org/abs/1206.5142}
  {arXiv:1206.5142 [hep-lat]} \BibitemShut {NoStop}%
%%CITATION = ARXIV:1206.5142;%%
\bibitem [{\citenamefont {Blum}\ \emph {et~al.}(2015)\citenamefont {Blum} \emph
  {et~al.}}]{Blum:2015ywa}%
  \BibitemOpen
  \bibfield  {author} {\bibinfo {author} {\bibfnamefont {T.}~\bibnamefont
  {Blum}} \emph {et~al.},\ }\href {\doibase 10.1103/PhysRevD.91.074502}
  {\bibfield  {journal} {\bibinfo  {journal} {Phys. Rev.}\ }\textbf {\bibinfo
  {volume} {D91}},\ \bibinfo {pages} {074502} (\bibinfo {year} {2015})},\
  \Eprint {http://arxiv.org/abs/1502.00263} {arXiv:1502.00263 [hep-lat]}
  \BibitemShut {NoStop}%
%%CITATION = ARXIV:1502.00263;%%
\bibitem [{\citenamefont {Bai}\ \emph {et~al.}(2015)\citenamefont {Bai} \emph
  {et~al.}}]{Bai:2015nea}%
  \BibitemOpen
  \bibfield  {author} {\bibinfo {author} {\bibfnamefont {Z.}~\bibnamefont
  {Bai}} \emph {et~al.} (\bibinfo {collaboration} {RBC, UKQCD}),\ }\href
  {\doibase 10.1103/PhysRevLett.115.212001} {\bibfield  {journal} {\bibinfo
  {journal} {Phys. Rev. Lett.}\ }\textbf {\bibinfo {volume} {115}},\ \bibinfo
  {pages} {212001} (\bibinfo {year} {2015})},\ \Eprint
  {http://arxiv.org/abs/1505.07863} {arXiv:1505.07863 [hep-lat]} \BibitemShut
  {NoStop}%
\bibitem [{\citenamefont {Abbott}\ \emph {et~al.}(2020)\citenamefont {Abbott}
  \emph {et~al.}}]{Abbott:2020hxn}%
  \BibitemOpen
  \bibfield  {author} {\bibinfo {author} {\bibfnamefont {R.}~\bibnamefont
  {Abbott}} \emph {et~al.},\ }\href@noop {} {\  (\bibinfo {year} {2020})},\
  \Eprint {http://arxiv.org/abs/2004.09440} {arXiv:2004.09440 [hep-lat]}
  \BibitemShut {NoStop}%
%%CITATION = ARXIV:2004.09440;%%
\bibitem [{\citenamefont {Briceno}\ \emph {et~al.}(2015)\citenamefont
  {Briceno}, \citenamefont {Dudek}, \citenamefont {Edwards}, \citenamefont
  {Shultz}, \citenamefont {Thomas},\ and\ \citenamefont
  {Wilson}}]{Briceno:2015dca}%
  \BibitemOpen
  \bibfield  {author} {\bibinfo {author} {\bibfnamefont {R.~A.}\ \bibnamefont
  {Briceno}}, \bibinfo {author} {\bibfnamefont {J.~J.}\ \bibnamefont {Dudek}},
  \bibinfo {author} {\bibfnamefont {R.~G.}\ \bibnamefont {Edwards}}, \bibinfo
  {author} {\bibfnamefont {C.~J.}\ \bibnamefont {Shultz}}, \bibinfo {author}
  {\bibfnamefont {C.~E.}\ \bibnamefont {Thomas}}, \ and\ \bibinfo {author}
  {\bibfnamefont {D.~J.}\ \bibnamefont {Wilson}},\ }\href {\doibase
  10.1103/PhysRevLett.115.242001} {\bibfield  {journal} {\bibinfo  {journal}
  {Phys. Rev. Lett.}\ }\textbf {\bibinfo {volume} {115}},\ \bibinfo {pages}
  {242001} (\bibinfo {year} {2015})},\ \Eprint
  {http://arxiv.org/abs/1507.06622} {arXiv:1507.06622 [hep-ph]} \BibitemShut
  {NoStop}%
%%CITATION = ARXIV:1507.06622;%%
\bibitem [{\citenamefont {Briceño}\ \emph {et~al.}(2016)\citenamefont
  {Briceño}, \citenamefont {Dudek}, \citenamefont {Edwards}, \citenamefont
  {Shultz}, \citenamefont {Thomas},\ and\ \citenamefont
  {Wilson}}]{Briceno:2016kkp}%
  \BibitemOpen
  \bibfield  {author} {\bibinfo {author} {\bibfnamefont {R.~A.}\ \bibnamefont
  {Briceño}}, \bibinfo {author} {\bibfnamefont {J.~J.}\ \bibnamefont {Dudek}},
  \bibinfo {author} {\bibfnamefont {R.~G.}\ \bibnamefont {Edwards}}, \bibinfo
  {author} {\bibfnamefont {C.~J.}\ \bibnamefont {Shultz}}, \bibinfo {author}
  {\bibfnamefont {C.~E.}\ \bibnamefont {Thomas}}, \ and\ \bibinfo {author}
  {\bibfnamefont {D.~J.}\ \bibnamefont {Wilson}},\ }\href {\doibase
  10.1103/PhysRevD.93.114508} {\bibfield  {journal} {\bibinfo  {journal} {Phys.
  Rev.}\ }\textbf {\bibinfo {volume} {D93}},\ \bibinfo {pages} {114508}
  (\bibinfo {year} {2016})},\ \Eprint {http://arxiv.org/abs/1604.03530}
  {arXiv:1604.03530 [hep-ph]} \BibitemShut {NoStop}%
%%CITATION = ARXIV:1604.03530;%%
\bibitem [{\citenamefont {Alexandrou}\ \emph {et~al.}(2018)\citenamefont
  {Alexandrou}, \citenamefont {Leskovec}, \citenamefont {Meinel}, \citenamefont
  {Negele}, \citenamefont {Paul}, \citenamefont {Petschlies}, \citenamefont
  {Pochinsky}, \citenamefont {Rendon},\ and\ \citenamefont
  {Syritsyn}}]{Alexandrou:2018jbt}%
  \BibitemOpen
  \bibfield  {author} {\bibinfo {author} {\bibfnamefont {C.}~\bibnamefont
  {Alexandrou}}, \bibinfo {author} {\bibfnamefont {L.}~\bibnamefont
  {Leskovec}}, \bibinfo {author} {\bibfnamefont {S.}~\bibnamefont {Meinel}},
  \bibinfo {author} {\bibfnamefont {J.}~\bibnamefont {Negele}}, \bibinfo
  {author} {\bibfnamefont {S.}~\bibnamefont {Paul}}, \bibinfo {author}
  {\bibfnamefont {M.}~\bibnamefont {Petschlies}}, \bibinfo {author}
  {\bibfnamefont {A.}~\bibnamefont {Pochinsky}}, \bibinfo {author}
  {\bibfnamefont {G.}~\bibnamefont {Rendon}}, \ and\ \bibinfo {author}
  {\bibfnamefont {S.}~\bibnamefont {Syritsyn}},\ }\href {\doibase
  10.1103/PhysRevD.98.074502} {\bibfield  {journal} {\bibinfo  {journal} {Phys.
  Rev.}\ }\textbf {\bibinfo {volume} {D98}},\ \bibinfo {pages} {074502}
  (\bibinfo {year} {2018})},\ \Eprint {http://arxiv.org/abs/1807.08357}
  {arXiv:1807.08357 [hep-lat]} \BibitemShut {NoStop}%
%%CITATION = ARXIV:1807.08357;%%
\bibitem [{\citenamefont {Baroni}\ \emph {et~al.}(2019)\citenamefont {Baroni},
  \citenamefont {Briceño}, \citenamefont {Hansen},\ and\ \citenamefont
  {Ortega-Gama}}]{Baroni:2018iau}%
  \BibitemOpen
  \bibfield  {author} {\bibinfo {author} {\bibfnamefont {A.}~\bibnamefont
  {Baroni}}, \bibinfo {author} {\bibfnamefont {R.~A.}\ \bibnamefont
  {Briceño}}, \bibinfo {author} {\bibfnamefont {M.~T.}\ \bibnamefont
  {Hansen}}, \ and\ \bibinfo {author} {\bibfnamefont {F.~G.}\ \bibnamefont
  {Ortega-Gama}},\ }\href {\doibase 10.1103/PhysRevD.100.034511} {\bibfield
  {journal} {\bibinfo  {journal} {Phys. Rev.}\ }\textbf {\bibinfo {volume}
  {D100}},\ \bibinfo {pages} {034511} (\bibinfo {year} {2019})},\ \Eprint
  {http://arxiv.org/abs/1812.10504} {arXiv:1812.10504 [hep-lat]} \BibitemShut
  {NoStop}%
%%CITATION = ARXIV:1812.10504;%%
\bibitem [{\citenamefont {Christ}(2010)}]{Christ:2010gi}%
  \BibitemOpen
  \bibfield  {author} {\bibinfo {author} {\bibfnamefont {N.~H.}\ \bibnamefont
  {Christ}} (\bibinfo {collaboration} {RBC, UKQCD}),\ }in\ \href@noop {} {\emph
  {\bibinfo {booktitle} {{Proceedings, 28th International Symposium on Lattice
  field theory (Lattice 2010)}: {Villasimius, Italy, June 14-19, 2010}}}}\
  (\bibinfo {year} {2010})\ \Eprint {http://arxiv.org/abs/1012.6034}
  {arXiv:1012.6034 [hep-lat]} \BibitemShut {NoStop}%
\bibitem [{\citenamefont {Christ}\ \emph {et~al.}(2014)\citenamefont {Christ},
  \citenamefont {Martinelli},\ and\ \citenamefont
  {Sachrajda}}]{Christ:2014qaa}%
  \BibitemOpen
  \bibfield  {author} {\bibinfo {author} {\bibfnamefont {N.~H.}\ \bibnamefont
  {Christ}}, \bibinfo {author} {\bibfnamefont {G.}~\bibnamefont {Martinelli}},
  \ and\ \bibinfo {author} {\bibfnamefont {C.~T.}\ \bibnamefont {Sachrajda}},\
  }\bibfield  {booktitle} {\emph {\bibinfo {booktitle} {{Proceedings, 31st
  International Symposium on Lattice Field Theory (Lattice 2013): Mainz,
  Germany, July 29-August 3, 2013}}},\ }\href {\doibase 10.22323/1.187.0399}
  {\bibfield  {journal} {\bibinfo  {journal} {PoS}\ }\textbf {\bibinfo {volume}
  {LATTICE2013}},\ \bibinfo {pages} {399} (\bibinfo {year} {2014})},\ \Eprint
  {http://arxiv.org/abs/1401.1362} {arXiv:1401.1362 [hep-lat]} \BibitemShut
  {NoStop}%
%%CITATION = ARXIV:1401.1362;%%
\bibitem [{\citenamefont {Christ}\ \emph
  {et~al.}(2015{\natexlab{a}})\citenamefont {Christ}, \citenamefont {Feng},
  \citenamefont {Martinelli},\ and\ \citenamefont
  {Sachrajda}}]{Christ:2015pwa}%
  \BibitemOpen
  \bibfield  {author} {\bibinfo {author} {\bibfnamefont {N.~H.}\ \bibnamefont
  {Christ}}, \bibinfo {author} {\bibfnamefont {X.}~\bibnamefont {Feng}},
  \bibinfo {author} {\bibfnamefont {G.}~\bibnamefont {Martinelli}}, \ and\
  \bibinfo {author} {\bibfnamefont {C.~T.}\ \bibnamefont {Sachrajda}},\ }\href
  {\doibase 10.1103/PhysRevD.91.114510} {\bibfield  {journal} {\bibinfo
  {journal} {Phys. Rev. D}\ }\textbf {\bibinfo {volume} {91}},\ \bibinfo
  {pages} {114510} (\bibinfo {year} {2015}{\natexlab{a}})},\ \Eprint
  {http://arxiv.org/abs/1504.01170} {arXiv:1504.01170 [hep-lat]} \BibitemShut
  {NoStop}%
\bibitem [{\citenamefont {Christ}\ \emph {et~al.}(2013)\citenamefont {Christ},
  \citenamefont {Izubuchi}, \citenamefont {Sachrajda}, \citenamefont {Soni},\
  and\ \citenamefont {Yu}}]{Christ:2012se}%
  \BibitemOpen
  \bibfield  {author} {\bibinfo {author} {\bibfnamefont {N.~H.}\ \bibnamefont
  {Christ}}, \bibinfo {author} {\bibfnamefont {T.}~\bibnamefont {Izubuchi}},
  \bibinfo {author} {\bibfnamefont {C.~T.}\ \bibnamefont {Sachrajda}}, \bibinfo
  {author} {\bibfnamefont {A.}~\bibnamefont {Soni}}, \ and\ \bibinfo {author}
  {\bibfnamefont {J.}~\bibnamefont {Yu}} (\bibinfo {collaboration} {RBC,
  UKQCD}),\ }\href {\doibase 10.1103/PhysRevD.88.014508} {\bibfield  {journal}
  {\bibinfo  {journal} {Phys. Rev.}\ }\textbf {\bibinfo {volume} {D88}},\
  \bibinfo {pages} {014508} (\bibinfo {year} {2013})},\ \Eprint
  {http://arxiv.org/abs/1212.5931} {arXiv:1212.5931 [hep-lat]} \BibitemShut
  {NoStop}%
%%CITATION = ARXIV:1212.5931;%%
\bibitem [{\citenamefont {Bai}\ \emph {et~al.}(2014)\citenamefont {Bai},
  \citenamefont {Christ}, \citenamefont {Izubuchi}, \citenamefont {Sachrajda},
  \citenamefont {Soni},\ and\ \citenamefont {Yu}}]{Bai:2014cva}%
  \BibitemOpen
  \bibfield  {author} {\bibinfo {author} {\bibfnamefont {Z.}~\bibnamefont
  {Bai}}, \bibinfo {author} {\bibfnamefont {N.~H.}\ \bibnamefont {Christ}},
  \bibinfo {author} {\bibfnamefont {T.}~\bibnamefont {Izubuchi}}, \bibinfo
  {author} {\bibfnamefont {C.~T.}\ \bibnamefont {Sachrajda}}, \bibinfo {author}
  {\bibfnamefont {A.}~\bibnamefont {Soni}}, \ and\ \bibinfo {author}
  {\bibfnamefont {J.}~\bibnamefont {Yu}},\ }\href {\doibase
  10.1103/PhysRevLett.113.112003} {\bibfield  {journal} {\bibinfo  {journal}
  {Phys. Rev. Lett.}\ }\textbf {\bibinfo {volume} {113}},\ \bibinfo {pages}
  {112003} (\bibinfo {year} {2014})},\ \Eprint {http://arxiv.org/abs/1406.0916}
  {arXiv:1406.0916 [hep-lat]} \BibitemShut {NoStop}%
%%CITATION = ARXIV:1406.0916;%%
\bibitem [{\citenamefont {Bai}(2017)}]{Bai:2016gzv}%
  \BibitemOpen
  \bibfield  {author} {\bibinfo {author} {\bibfnamefont {Z.}~\bibnamefont
  {Bai}},\ }\bibfield  {booktitle} {\emph {\bibinfo {booktitle} {{Proceedings,
  34th International Symposium on Lattice Field Theory (Lattice 2016):
  Southampton, UK, July 24-30, 2016}}},\ }\href {\doibase 10.22323/1.256.0309}
  {\bibfield  {journal} {\bibinfo  {journal} {PoS}\ }\textbf {\bibinfo {volume}
  {LATTICE2016}},\ \bibinfo {pages} {309} (\bibinfo {year} {2017})},\ \Eprint
  {http://arxiv.org/abs/1611.06601} {arXiv:1611.06601 [hep-lat]} \BibitemShut
  {NoStop}%
%%CITATION = ARXIV:1611.06601;%%
\bibitem [{\citenamefont {Wang}(2020)}]{Wang:2020jpi}%
  \BibitemOpen
  \bibfield  {author} {\bibinfo {author} {\bibfnamefont {B.}~\bibnamefont
  {Wang}},\ }\href@noop {} {\  (\bibinfo {year} {2020})},\ \Eprint
  {http://arxiv.org/abs/2001.06374} {arXiv:2001.06374 [hep-lat]} \BibitemShut
  {NoStop}%
%%CITATION = ARXIV:2001.06374;%%
\bibitem [{\citenamefont {Christ}\ \emph
  {et~al.}(2015{\natexlab{b}})\citenamefont {Christ}, \citenamefont {Feng},
  \citenamefont {Portelli},\ and\ \citenamefont {Sachrajda}}]{Christ:2015aha}%
  \BibitemOpen
  \bibfield  {author} {\bibinfo {author} {\bibfnamefont {N.~H.}\ \bibnamefont
  {Christ}}, \bibinfo {author} {\bibfnamefont {X.}~\bibnamefont {Feng}},
  \bibinfo {author} {\bibfnamefont {A.}~\bibnamefont {Portelli}}, \ and\
  \bibinfo {author} {\bibfnamefont {C.~T.}\ \bibnamefont {Sachrajda}} (\bibinfo
  {collaboration} {RBC, UKQCD}),\ }\href {\doibase 10.1103/PhysRevD.92.094512}
  {\bibfield  {journal} {\bibinfo  {journal} {Phys. Rev.}\ }\textbf {\bibinfo
  {volume} {D92}},\ \bibinfo {pages} {094512} (\bibinfo {year}
  {2015}{\natexlab{b}})},\ \Eprint {http://arxiv.org/abs/1507.03094}
  {arXiv:1507.03094 [hep-lat]} \BibitemShut {NoStop}%
%%CITATION = ARXIV:1507.03094;%%
\bibitem [{\citenamefont {Christ}\ \emph
  {et~al.}(2016{\natexlab{a}})\citenamefont {Christ}, \citenamefont {Feng},
  \citenamefont {Portelli},\ and\ \citenamefont {Sachrajda}}]{Christ:2016eae}%
  \BibitemOpen
  \bibfield  {author} {\bibinfo {author} {\bibfnamefont {N.~H.}\ \bibnamefont
  {Christ}}, \bibinfo {author} {\bibfnamefont {X.}~\bibnamefont {Feng}},
  \bibinfo {author} {\bibfnamefont {A.}~\bibnamefont {Portelli}}, \ and\
  \bibinfo {author} {\bibfnamefont {C.~T.}\ \bibnamefont {Sachrajda}} (\bibinfo
  {collaboration} {RBC, UKQCD}),\ }\href {\doibase 10.1103/PhysRevD.93.114517}
  {\bibfield  {journal} {\bibinfo  {journal} {Phys. Rev.}\ }\textbf {\bibinfo
  {volume} {D93}},\ \bibinfo {pages} {114517} (\bibinfo {year}
  {2016}{\natexlab{a}})},\ \Eprint {http://arxiv.org/abs/1605.04442}
  {arXiv:1605.04442 [hep-lat]} \BibitemShut {NoStop}%
%%CITATION = ARXIV:1605.04442;%%
\bibitem [{\citenamefont {Christ}\ \emph
  {et~al.}(2016{\natexlab{b}})\citenamefont {Christ}, \citenamefont {Feng},
  \citenamefont {Juttner}, \citenamefont {Lawson}, \citenamefont {Portelli},\
  and\ \citenamefont {Sachrajda}}]{Christ:2016mmq}%
  \BibitemOpen
  \bibfield  {author} {\bibinfo {author} {\bibfnamefont {N.~H.}\ \bibnamefont
  {Christ}}, \bibinfo {author} {\bibfnamefont {X.}~\bibnamefont {Feng}},
  \bibinfo {author} {\bibfnamefont {A.}~\bibnamefont {Juttner}}, \bibinfo
  {author} {\bibfnamefont {A.}~\bibnamefont {Lawson}}, \bibinfo {author}
  {\bibfnamefont {A.}~\bibnamefont {Portelli}}, \ and\ \bibinfo {author}
  {\bibfnamefont {C.~T.}\ \bibnamefont {Sachrajda}},\ }\href {\doibase
  10.1103/PhysRevD.94.114516} {\bibfield  {journal} {\bibinfo  {journal} {Phys.
  Rev.}\ }\textbf {\bibinfo {volume} {D94}},\ \bibinfo {pages} {114516}
  (\bibinfo {year} {2016}{\natexlab{b}})},\ \Eprint
  {http://arxiv.org/abs/1608.07585} {arXiv:1608.07585 [hep-lat]} \BibitemShut
  {NoStop}%
%%CITATION = ARXIV:1608.07585;%%
\bibitem [{\citenamefont {Bai}\ \emph {et~al.}(2017)\citenamefont {Bai},
  \citenamefont {Christ}, \citenamefont {Feng}, \citenamefont {Lawson},
  \citenamefont {Portelli},\ and\ \citenamefont {Sachrajda}}]{Bai:2017fkh}%
  \BibitemOpen
  \bibfield  {author} {\bibinfo {author} {\bibfnamefont {Z.}~\bibnamefont
  {Bai}}, \bibinfo {author} {\bibfnamefont {N.~H.}\ \bibnamefont {Christ}},
  \bibinfo {author} {\bibfnamefont {X.}~\bibnamefont {Feng}}, \bibinfo {author}
  {\bibfnamefont {A.}~\bibnamefont {Lawson}}, \bibinfo {author} {\bibfnamefont
  {A.}~\bibnamefont {Portelli}}, \ and\ \bibinfo {author} {\bibfnamefont
  {C.~T.}\ \bibnamefont {Sachrajda}},\ }\href {\doibase
  10.1103/PhysRevLett.118.252001} {\bibfield  {journal} {\bibinfo  {journal}
  {Phys. Rev. Lett.}\ }\textbf {\bibinfo {volume} {118}},\ \bibinfo {pages}
  {252001} (\bibinfo {year} {2017})},\ \Eprint
  {http://arxiv.org/abs/1701.02858} {arXiv:1701.02858 [hep-lat]} \BibitemShut
  {NoStop}%
%%CITATION = ARXIV:1701.02858;%%
\bibitem [{\citenamefont {Bai}\ \emph {et~al.}(2018)\citenamefont {Bai},
  \citenamefont {Christ}, \citenamefont {Feng}, \citenamefont {Lawson},
  \citenamefont {Portelli},\ and\ \citenamefont {Sachrajda}}]{Bai:2018hqu}%
  \BibitemOpen
  \bibfield  {author} {\bibinfo {author} {\bibfnamefont {Z.}~\bibnamefont
  {Bai}}, \bibinfo {author} {\bibfnamefont {N.~H.}\ \bibnamefont {Christ}},
  \bibinfo {author} {\bibfnamefont {X.}~\bibnamefont {Feng}}, \bibinfo {author}
  {\bibfnamefont {A.}~\bibnamefont {Lawson}}, \bibinfo {author} {\bibfnamefont
  {A.}~\bibnamefont {Portelli}}, \ and\ \bibinfo {author} {\bibfnamefont
  {C.~T.}\ \bibnamefont {Sachrajda}},\ }\href {\doibase
  10.1103/PhysRevD.98.074509} {\bibfield  {journal} {\bibinfo  {journal} {Phys.
  Rev.}\ }\textbf {\bibinfo {volume} {D98}},\ \bibinfo {pages} {074509}
  (\bibinfo {year} {2018})},\ \Eprint {http://arxiv.org/abs/1806.11520}
  {arXiv:1806.11520 [hep-lat]} \BibitemShut {NoStop}%
%%CITATION = ARXIV:1806.11520;%%
\bibitem [{\citenamefont {Christ}\ \emph {et~al.}(2019)\citenamefont {Christ},
  \citenamefont {Feng}, \citenamefont {Portelli},\ and\ \citenamefont
  {Sachrajda}}]{Christ:2019dxu}%
  \BibitemOpen
  \bibfield  {author} {\bibinfo {author} {\bibfnamefont {N.~H.}\ \bibnamefont
  {Christ}}, \bibinfo {author} {\bibfnamefont {X.}~\bibnamefont {Feng}},
  \bibinfo {author} {\bibfnamefont {A.}~\bibnamefont {Portelli}}, \ and\
  \bibinfo {author} {\bibfnamefont {C.~T.}\ \bibnamefont {Sachrajda}} (\bibinfo
  {collaboration} {RBC, UKQCD}),\ }\href {\doibase 10.1103/PhysRevD.100.114506}
  {\bibfield  {journal} {\bibinfo  {journal} {Phys. Rev.}\ }\textbf {\bibinfo
  {volume} {D100}},\ \bibinfo {pages} {114506} (\bibinfo {year} {2019})},\
  \Eprint {http://arxiv.org/abs/1910.10644} {arXiv:1910.10644 [hep-lat]}
  \BibitemShut {NoStop}%
%%CITATION = ARXIV:1910.10644;%%
\bibitem [{\citenamefont {Briceño}\ \emph {et~al.}(2020)\citenamefont
  {Briceño}, \citenamefont {Davoudi}, \citenamefont {Hansen}, \citenamefont
  {Schindler},\ and\ \citenamefont {Baroni}}]{Briceno:2019opb}%
  \BibitemOpen
  \bibfield  {author} {\bibinfo {author} {\bibfnamefont {R.~A.}\ \bibnamefont
  {Briceño}}, \bibinfo {author} {\bibfnamefont {Z.}~\bibnamefont {Davoudi}},
  \bibinfo {author} {\bibfnamefont {M.~T.}\ \bibnamefont {Hansen}}, \bibinfo
  {author} {\bibfnamefont {M.~R.}\ \bibnamefont {Schindler}}, \ and\ \bibinfo
  {author} {\bibfnamefont {A.}~\bibnamefont {Baroni}},\ }\href {\doibase
  10.1103/PhysRevD.101.014509} {\bibfield  {journal} {\bibinfo  {journal}
  {Phys. Rev. D}\ }\textbf {\bibinfo {volume} {101}},\ \bibinfo {pages}
  {014509} (\bibinfo {year} {2020})},\ \Eprint
  {http://arxiv.org/abs/1911.04036} {arXiv:1911.04036 [hep-lat]} \BibitemShut
  {NoStop}%
\bibitem [{\citenamefont {Kim}\ \emph {et~al.}(2005)\citenamefont {Kim},
  \citenamefont {Sachrajda},\ and\ \citenamefont {Sharpe}}]{Kim:2005gf}%
  \BibitemOpen
  \bibfield  {author} {\bibinfo {author} {\bibfnamefont {C.}~\bibnamefont
  {Kim}}, \bibinfo {author} {\bibfnamefont {C.}~\bibnamefont {Sachrajda}}, \
  and\ \bibinfo {author} {\bibfnamefont {S.~R.}\ \bibnamefont {Sharpe}},\
  }\href {\doibase 10.1016/j.nuclphysb.2005.08.029} {\bibfield  {journal}
  {\bibinfo  {journal} {Nucl. Phys. B}\ }\textbf {\bibinfo {volume} {727}},\
  \bibinfo {pages} {218} (\bibinfo {year} {2005})},\ \Eprint
  {http://arxiv.org/abs/hep-lat/0507006} {arXiv:hep-lat/0507006} \BibitemShut
  {NoStop}%
\bibitem [{\citenamefont {He}\ \emph {et~al.}(2005)\citenamefont {He},
  \citenamefont {Feng},\ and\ \citenamefont {Liu}}]{He:2005ey}%
  \BibitemOpen
  \bibfield  {author} {\bibinfo {author} {\bibfnamefont {S.}~\bibnamefont
  {He}}, \bibinfo {author} {\bibfnamefont {X.}~\bibnamefont {Feng}}, \ and\
  \bibinfo {author} {\bibfnamefont {C.}~\bibnamefont {Liu}},\ }\href {\doibase
  10.1088/1126-6708/2005/07/011} {\bibfield  {journal} {\bibinfo  {journal}
  {JHEP}\ }\textbf {\bibinfo {volume} {07}},\ \bibinfo {pages} {011} (\bibinfo
  {year} {2005})},\ \Eprint {http://arxiv.org/abs/hep-lat/0504019}
  {arXiv:hep-lat/0504019} \BibitemShut {NoStop}%
\bibitem [{\citenamefont {Liu}\ \emph {et~al.}(2006)\citenamefont {Liu},
  \citenamefont {Feng},\ and\ \citenamefont {He}}]{Liu:2005kr}%
  \BibitemOpen
  \bibfield  {author} {\bibinfo {author} {\bibfnamefont {C.}~\bibnamefont
  {Liu}}, \bibinfo {author} {\bibfnamefont {X.}~\bibnamefont {Feng}}, \ and\
  \bibinfo {author} {\bibfnamefont {S.}~\bibnamefont {He}},\ }\href {\doibase
  10.1142/S0217751X06032150} {\bibfield  {journal} {\bibinfo  {journal} {Int.\
  J.\ Mod.\ Phys.\ A}\ }\textbf {\bibinfo {volume} {21}},\ \bibinfo {pages}
  {847} (\bibinfo {year} {2006})},\ \Eprint
  {http://arxiv.org/abs/hep-lat/0508022} {arXiv:hep-lat/0508022} \BibitemShut
  {NoStop}%
\bibitem [{\citenamefont {Christ}\ and\ \citenamefont
  {Feng}(2018)}]{Christ:2017pze}%
  \BibitemOpen
  \bibfield  {author} {\bibinfo {author} {\bibfnamefont {N.}~\bibnamefont
  {Christ}}\ and\ \bibinfo {author} {\bibfnamefont {X.}~\bibnamefont {Feng}},\
  }\href {\doibase 10.1051/epjconf/201817513016} {\bibfield  {journal}
  {\bibinfo  {journal} {EPJ Web Conf.}\ }\textbf {\bibinfo {volume} {175}},\
  \bibinfo {pages} {13016} (\bibinfo {year} {2018})},\ \Eprint
  {http://arxiv.org/abs/1711.09339} {arXiv:1711.09339 [hep-lat]} \BibitemShut
  {NoStop}%
\bibitem [{\citenamefont {Hansen}\ and\ \citenamefont
  {Sharpe}(2012)}]{Hansen:2012tf}%
  \BibitemOpen
  \bibfield  {author} {\bibinfo {author} {\bibfnamefont {M.~T.}\ \bibnamefont
  {Hansen}}\ and\ \bibinfo {author} {\bibfnamefont {S.~R.}\ \bibnamefont
  {Sharpe}},\ }\href {\doibase 10.1103/PhysRevD.86.016007} {\bibfield
  {journal} {\bibinfo  {journal} {Phys.\ Rev.\ D}\ }\textbf {\bibinfo {volume}
  {86}},\ \bibinfo {pages} {016007} (\bibinfo {year} {2012})},\ \Eprint
  {http://arxiv.org/abs/1204.0826} {arXiv:1204.0826 [hep-lat]} \BibitemShut
  {NoStop}%
\bibitem [{\citenamefont {Lage}\ \emph {et~al.}(2009)\citenamefont {Lage},
  \citenamefont {Meissner},\ and\ \citenamefont {Rusetsky}}]{Lage:2009zv}%
  \BibitemOpen
  \bibfield  {author} {\bibinfo {author} {\bibfnamefont {M.}~\bibnamefont
  {Lage}}, \bibinfo {author} {\bibfnamefont {U.-G.}\ \bibnamefont {Meissner}},
  \ and\ \bibinfo {author} {\bibfnamefont {A.}~\bibnamefont {Rusetsky}},\
  }\href {\doibase 10.1016/j.physletb.2009.10.055} {\bibfield  {journal}
  {\bibinfo  {journal} {Phys. Lett.}\ }\textbf {\bibinfo {volume} {B681}},\
  \bibinfo {pages} {439} (\bibinfo {year} {2009})},\ \Eprint
  {http://arxiv.org/abs/0905.0069} {arXiv:0905.0069 [hep-lat]} \BibitemShut
  {NoStop}%
%%CITATION = ARXIV:0905.0069;%%
\bibitem [{\citenamefont {Bernard}\ \emph {et~al.}(2011)\citenamefont
  {Bernard}, \citenamefont {Lage}, \citenamefont {Meissner},\ and\
  \citenamefont {Rusetsky}}]{Bernard:2010fp}%
  \BibitemOpen
  \bibfield  {author} {\bibinfo {author} {\bibfnamefont {V.}~\bibnamefont
  {Bernard}}, \bibinfo {author} {\bibfnamefont {M.}~\bibnamefont {Lage}},
  \bibinfo {author} {\bibfnamefont {U.~G.}\ \bibnamefont {Meissner}}, \ and\
  \bibinfo {author} {\bibfnamefont {A.}~\bibnamefont {Rusetsky}},\ }\href
  {\doibase 10.1007/JHEP01(2011)019} {\bibfield  {journal} {\bibinfo  {journal}
  {JHEP}\ }\textbf {\bibinfo {volume} {01}},\ \bibinfo {pages} {019} (\bibinfo
  {year} {2011})},\ \Eprint {http://arxiv.org/abs/1010.6018} {arXiv:1010.6018
  [hep-lat]} \BibitemShut {NoStop}%
%%CITATION = ARXIV:1010.6018;%%
\bibitem [{\citenamefont {Doring}\ \emph {et~al.}(2011)\citenamefont {Doring},
  \citenamefont {Meissner}, \citenamefont {Oset},\ and\ \citenamefont
  {Rusetsky}}]{Doring:2011vk}%
  \BibitemOpen
  \bibfield  {author} {\bibinfo {author} {\bibfnamefont {M.}~\bibnamefont
  {Doring}}, \bibinfo {author} {\bibfnamefont {U.-G.}\ \bibnamefont
  {Meissner}}, \bibinfo {author} {\bibfnamefont {E.}~\bibnamefont {Oset}}, \
  and\ \bibinfo {author} {\bibfnamefont {A.}~\bibnamefont {Rusetsky}},\ }\href
  {\doibase 10.1140/epja/i2011-11139-7} {\bibfield  {journal} {\bibinfo
  {journal} {Eur. Phys. J.}\ }\textbf {\bibinfo {volume} {A47}},\ \bibinfo
  {pages} {139} (\bibinfo {year} {2011})},\ \Eprint
  {http://arxiv.org/abs/1107.3988} {arXiv:1107.3988 [hep-lat]} \BibitemShut
  {NoStop}%
%%CITATION = ARXIV:1107.3988;%%
\bibitem [{\citenamefont {Aoki}\ \emph
  {et~al.}(2011{\natexlab{b}})\citenamefont {Aoki}, \citenamefont {Ishii},
  \citenamefont {Doi}, \citenamefont {Hatsuda}, \citenamefont {Ikeda},
  \citenamefont {Inoue}, \citenamefont {Murano}, \citenamefont {Nemura},\ and\
  \citenamefont {Sasaki}}]{Aoki:2011gt}%
  \BibitemOpen
  \bibfield  {author} {\bibinfo {author} {\bibfnamefont {S.}~\bibnamefont
  {Aoki}}, \bibinfo {author} {\bibfnamefont {N.}~\bibnamefont {Ishii}},
  \bibinfo {author} {\bibfnamefont {T.}~\bibnamefont {Doi}}, \bibinfo {author}
  {\bibfnamefont {T.}~\bibnamefont {Hatsuda}}, \bibinfo {author} {\bibfnamefont
  {Y.}~\bibnamefont {Ikeda}}, \bibinfo {author} {\bibfnamefont
  {T.}~\bibnamefont {Inoue}}, \bibinfo {author} {\bibfnamefont
  {K.}~\bibnamefont {Murano}}, \bibinfo {author} {\bibfnamefont
  {H.}~\bibnamefont {Nemura}}, \ and\ \bibinfo {author} {\bibfnamefont
  {K.}~\bibnamefont {Sasaki}} (\bibinfo {collaboration} {HAL QCD}),\ }\href
  {\doibase 10.2183/pjab.87.509} {\bibfield  {journal} {\bibinfo  {journal}
  {Proc. Japan Acad.}\ }\textbf {\bibinfo {volume} {B87}},\ \bibinfo {pages}
  {509} (\bibinfo {year} {2011}{\natexlab{b}})},\ \Eprint
  {http://arxiv.org/abs/1106.2281} {arXiv:1106.2281 [hep-lat]} \BibitemShut
  {NoStop}%
%%CITATION = ARXIV:1106.2281;%%
\bibitem [{\citenamefont {Beane}\ \emph {et~al.}(2004)\citenamefont {Beane},
  \citenamefont {Bedaque}, \citenamefont {Parreno},\ and\ \citenamefont
  {Savage}}]{Beane:2003da}%
  \BibitemOpen
  \bibfield  {author} {\bibinfo {author} {\bibfnamefont {S.}~\bibnamefont
  {Beane}}, \bibinfo {author} {\bibfnamefont {P.}~\bibnamefont {Bedaque}},
  \bibinfo {author} {\bibfnamefont {A.}~\bibnamefont {Parreno}}, \ and\
  \bibinfo {author} {\bibfnamefont {M.}~\bibnamefont {Savage}},\ }\href
  {\doibase 10.1016/j.physletb.2004.02.007} {\bibfield  {journal} {\bibinfo
  {journal} {Phys. Lett. B}\ }\textbf {\bibinfo {volume} {585}},\ \bibinfo
  {pages} {106} (\bibinfo {year} {2004})},\ \Eprint
  {http://arxiv.org/abs/hep-lat/0312004} {arXiv:hep-lat/0312004} \BibitemShut
  {NoStop}%
\bibitem [{\citenamefont {Sasaki}\ and\ \citenamefont
  {Yamazaki}(2006)}]{Sasaki:2006jn}%
  \BibitemOpen
  \bibfield  {author} {\bibinfo {author} {\bibfnamefont {S.}~\bibnamefont
  {Sasaki}}\ and\ \bibinfo {author} {\bibfnamefont {T.}~\bibnamefont
  {Yamazaki}},\ }\href {\doibase 10.1103/PhysRevD.74.114507} {\bibfield
  {journal} {\bibinfo  {journal} {Phys. Rev. D}\ }\textbf {\bibinfo {volume}
  {74}},\ \bibinfo {pages} {114507} (\bibinfo {year} {2006})},\ \Eprint
  {http://arxiv.org/abs/hep-lat/0610081} {arXiv:hep-lat/0610081} \BibitemShut
  {NoStop}%
\bibitem [{\citenamefont {Engel}\ and\ \citenamefont
  {Menéndez}(2017)}]{Engel:2016xgb}%
  \BibitemOpen
  \bibfield  {author} {\bibinfo {author} {\bibfnamefont {J.}~\bibnamefont
  {Engel}}\ and\ \bibinfo {author} {\bibfnamefont {J.}~\bibnamefont
  {Menéndez}},\ }\href {\doibase 10.1088/1361-6633/aa5bc5} {\bibfield
  {journal} {\bibinfo  {journal} {Rept. Prog. Phys.}\ }\textbf {\bibinfo
  {volume} {80}},\ \bibinfo {pages} {046301} (\bibinfo {year} {2017})},\
  \Eprint {http://arxiv.org/abs/1610.06548} {arXiv:1610.06548 [nucl-th]}
  \BibitemShut {NoStop}%
%%CITATION = ARXIV:1610.06548;%%
\bibitem [{\citenamefont {Cirigliano}\ \emph
  {et~al.}(2017{\natexlab{a}})\citenamefont {Cirigliano}, \citenamefont
  {Dekens}, \citenamefont {Graesser},\ and\ \citenamefont
  {Mereghetti}}]{Cirigliano:2017ymo}%
  \BibitemOpen
  \bibfield  {author} {\bibinfo {author} {\bibfnamefont {V.}~\bibnamefont
  {Cirigliano}}, \bibinfo {author} {\bibfnamefont {W.}~\bibnamefont {Dekens}},
  \bibinfo {author} {\bibfnamefont {M.}~\bibnamefont {Graesser}}, \ and\
  \bibinfo {author} {\bibfnamefont {E.}~\bibnamefont {Mereghetti}},\ }\href
  {\doibase 10.1016/j.physletb.2017.04.020} {\bibfield  {journal} {\bibinfo
  {journal} {Phys. Lett. B}\ }\textbf {\bibinfo {volume} {769}},\ \bibinfo
  {pages} {460} (\bibinfo {year} {2017}{\natexlab{a}})},\ \Eprint
  {http://arxiv.org/abs/1701.01443} {arXiv:1701.01443 [hep-ph]} \BibitemShut
  {NoStop}%
\bibitem [{\citenamefont {Cirigliano}\ \emph
  {et~al.}(2017{\natexlab{b}})\citenamefont {Cirigliano}, \citenamefont
  {Dekens}, \citenamefont {de~Vries}, \citenamefont {Graesser},\ and\
  \citenamefont {Mereghetti}}]{Cirigliano:2017djv}%
  \BibitemOpen
  \bibfield  {author} {\bibinfo {author} {\bibfnamefont {V.}~\bibnamefont
  {Cirigliano}}, \bibinfo {author} {\bibfnamefont {W.}~\bibnamefont {Dekens}},
  \bibinfo {author} {\bibfnamefont {J.}~\bibnamefont {de~Vries}}, \bibinfo
  {author} {\bibfnamefont {M.}~\bibnamefont {Graesser}}, \ and\ \bibinfo
  {author} {\bibfnamefont {E.}~\bibnamefont {Mereghetti}},\ }\href {\doibase
  10.1007/JHEP12(2017)082} {\bibfield  {journal} {\bibinfo  {journal} {JHEP}\
  }\textbf {\bibinfo {volume} {12}},\ \bibinfo {pages} {082} (\bibinfo {year}
  {2017}{\natexlab{b}})},\ \Eprint {http://arxiv.org/abs/1708.09390}
  {arXiv:1708.09390 [hep-ph]} \BibitemShut {NoStop}%
\bibitem [{\citenamefont {Cirigliano}\ \emph
  {et~al.}(2018{\natexlab{a}})\citenamefont {Cirigliano}, \citenamefont
  {Dekens}, \citenamefont {Mereghetti},\ and\ \citenamefont
  {Walker-Loud}}]{Cirigliano:2017tvr}%
  \BibitemOpen
  \bibfield  {author} {\bibinfo {author} {\bibfnamefont {V.}~\bibnamefont
  {Cirigliano}}, \bibinfo {author} {\bibfnamefont {W.}~\bibnamefont {Dekens}},
  \bibinfo {author} {\bibfnamefont {E.}~\bibnamefont {Mereghetti}}, \ and\
  \bibinfo {author} {\bibfnamefont {A.}~\bibnamefont {Walker-Loud}},\ }\href
  {\doibase 10.1103/PhysRevC.97.065501} {\bibfield  {journal} {\bibinfo
  {journal} {Phys. Rev. C}\ }\textbf {\bibinfo {volume} {97}},\ \bibinfo
  {pages} {065501} (\bibinfo {year} {2018}{\natexlab{a}})},\ \bibinfo {note}
  {[Erratum: Phys.Rev.C 100, 019903 (2019)]},\ \Eprint
  {http://arxiv.org/abs/1710.01729} {arXiv:1710.01729 [hep-ph]} \BibitemShut
  {NoStop}%
\bibitem [{\citenamefont {Pastore}\ \emph {et~al.}(2018)\citenamefont
  {Pastore}, \citenamefont {Carlson}, \citenamefont {Cirigliano}, \citenamefont
  {Dekens}, \citenamefont {Mereghetti},\ and\ \citenamefont
  {Wiringa}}]{Pastore:2017ofx}%
  \BibitemOpen
  \bibfield  {author} {\bibinfo {author} {\bibfnamefont {S.}~\bibnamefont
  {Pastore}}, \bibinfo {author} {\bibfnamefont {J.}~\bibnamefont {Carlson}},
  \bibinfo {author} {\bibfnamefont {V.}~\bibnamefont {Cirigliano}}, \bibinfo
  {author} {\bibfnamefont {W.}~\bibnamefont {Dekens}}, \bibinfo {author}
  {\bibfnamefont {E.}~\bibnamefont {Mereghetti}}, \ and\ \bibinfo {author}
  {\bibfnamefont {R.}~\bibnamefont {Wiringa}},\ }\href {\doibase
  10.1103/PhysRevC.97.014606} {\bibfield  {journal} {\bibinfo  {journal} {Phys.
  Rev. C}\ }\textbf {\bibinfo {volume} {97}},\ \bibinfo {pages} {014606}
  (\bibinfo {year} {2018})},\ \Eprint {http://arxiv.org/abs/1710.05026}
  {arXiv:1710.05026 [nucl-th]} \BibitemShut {NoStop}%
\bibitem [{\citenamefont {Cirigliano}\ \emph
  {et~al.}(2018{\natexlab{b}})\citenamefont {Cirigliano}, \citenamefont
  {Dekens}, \citenamefont {De~Vries}, \citenamefont {Graesser}, \citenamefont
  {Mereghetti}, \citenamefont {Pastore},\ and\ \citenamefont
  {Van~Kolck}}]{Cirigliano:2018hja}%
  \BibitemOpen
  \bibfield  {author} {\bibinfo {author} {\bibfnamefont {V.}~\bibnamefont
  {Cirigliano}}, \bibinfo {author} {\bibfnamefont {W.}~\bibnamefont {Dekens}},
  \bibinfo {author} {\bibfnamefont {J.}~\bibnamefont {De~Vries}}, \bibinfo
  {author} {\bibfnamefont {M.~L.}\ \bibnamefont {Graesser}}, \bibinfo {author}
  {\bibfnamefont {E.}~\bibnamefont {Mereghetti}}, \bibinfo {author}
  {\bibfnamefont {S.}~\bibnamefont {Pastore}}, \ and\ \bibinfo {author}
  {\bibfnamefont {U.}~\bibnamefont {Van~Kolck}},\ }\href {\doibase
  10.1103/PhysRevLett.120.202001} {\bibfield  {journal} {\bibinfo  {journal}
  {Phys. Rev. Lett.}\ }\textbf {\bibinfo {volume} {120}},\ \bibinfo {pages}
  {202001} (\bibinfo {year} {2018}{\natexlab{b}})},\ \Eprint
  {http://arxiv.org/abs/1802.10097} {arXiv:1802.10097 [hep-ph]} \BibitemShut
  {NoStop}%
\bibitem [{\citenamefont {Cirigliano}\ \emph
  {et~al.}(2018{\natexlab{c}})\citenamefont {Cirigliano}, \citenamefont
  {Dekens}, \citenamefont {de~Vries}, \citenamefont {Graesser},\ and\
  \citenamefont {Mereghetti}}]{Cirigliano:2018yza}%
  \BibitemOpen
  \bibfield  {author} {\bibinfo {author} {\bibfnamefont {V.}~\bibnamefont
  {Cirigliano}}, \bibinfo {author} {\bibfnamefont {W.}~\bibnamefont {Dekens}},
  \bibinfo {author} {\bibfnamefont {J.}~\bibnamefont {de~Vries}}, \bibinfo
  {author} {\bibfnamefont {M.}~\bibnamefont {Graesser}}, \ and\ \bibinfo
  {author} {\bibfnamefont {E.}~\bibnamefont {Mereghetti}},\ }\href {\doibase
  10.1007/JHEP12(2018)097} {\bibfield  {journal} {\bibinfo  {journal} {JHEP}\
  }\textbf {\bibinfo {volume} {12}},\ \bibinfo {pages} {097} (\bibinfo {year}
  {2018}{\natexlab{c}})},\ \Eprint {http://arxiv.org/abs/1806.02780}
  {arXiv:1806.02780 [hep-ph]} \BibitemShut {NoStop}%
\bibitem [{\citenamefont {Cirigliano}\ \emph {et~al.}(2019)\citenamefont
  {Cirigliano}, \citenamefont {Dekens}, \citenamefont {De~Vries}, \citenamefont
  {Graesser}, \citenamefont {Mereghetti}, \citenamefont {Pastore},
  \citenamefont {Piarulli}, \citenamefont {Van~Kolck},\ and\ \citenamefont
  {Wiringa}}]{Cirigliano:2019vdj}%
  \BibitemOpen
  \bibfield  {author} {\bibinfo {author} {\bibfnamefont {V.}~\bibnamefont
  {Cirigliano}}, \bibinfo {author} {\bibfnamefont {W.}~\bibnamefont {Dekens}},
  \bibinfo {author} {\bibfnamefont {J.}~\bibnamefont {De~Vries}}, \bibinfo
  {author} {\bibfnamefont {M.}~\bibnamefont {Graesser}}, \bibinfo {author}
  {\bibfnamefont {E.}~\bibnamefont {Mereghetti}}, \bibinfo {author}
  {\bibfnamefont {S.}~\bibnamefont {Pastore}}, \bibinfo {author} {\bibfnamefont
  {M.}~\bibnamefont {Piarulli}}, \bibinfo {author} {\bibfnamefont
  {U.}~\bibnamefont {Van~Kolck}}, \ and\ \bibinfo {author} {\bibfnamefont
  {R.}~\bibnamefont {Wiringa}},\ }\href {\doibase 10.1103/PhysRevC.100.055504}
  {\bibfield  {journal} {\bibinfo  {journal} {Phys. Rev. C}\ }\textbf {\bibinfo
  {volume} {100}},\ \bibinfo {pages} {055504} (\bibinfo {year} {2019})},\
  \Eprint {http://arxiv.org/abs/1907.11254} {arXiv:1907.11254 [nucl-th]}
  \BibitemShut {NoStop}%
\bibitem [{\citenamefont {Dekens}\ \emph {et~al.}(2020)\citenamefont {Dekens},
  \citenamefont {de~Vries}, \citenamefont {Fuyuto}, \citenamefont
  {Mereghetti},\ and\ \citenamefont {Zhou}}]{Dekens:2020ttz}%
  \BibitemOpen
  \bibfield  {author} {\bibinfo {author} {\bibfnamefont {W.}~\bibnamefont
  {Dekens}}, \bibinfo {author} {\bibfnamefont {J.}~\bibnamefont {de~Vries}},
  \bibinfo {author} {\bibfnamefont {K.}~\bibnamefont {Fuyuto}}, \bibinfo
  {author} {\bibfnamefont {E.}~\bibnamefont {Mereghetti}}, \ and\ \bibinfo
  {author} {\bibfnamefont {G.}~\bibnamefont {Zhou}},\ }\href@noop {} {\
  (\bibinfo {year} {2020})},\ \Eprint {http://arxiv.org/abs/2002.07182}
  {arXiv:2002.07182 [hep-ph]} \BibitemShut {NoStop}%
\bibitem [{\citenamefont {Nicholson}\ \emph {et~al.}(2018)\citenamefont
  {Nicholson} \emph {et~al.}}]{Nicholson:2018mwc}%
  \BibitemOpen
  \bibfield  {author} {\bibinfo {author} {\bibfnamefont {A.}~\bibnamefont
  {Nicholson}} \emph {et~al.},\ }\href@noop {} {\  (\bibinfo {year} {2018})},\
  \Eprint {http://arxiv.org/abs/1805.02634} {arXiv:1805.02634 [nucl-th]}
  \BibitemShut {NoStop}%
%%CITATION = ARXIV:1805.02634;%%
\bibitem [{\citenamefont {Feng}\ \emph {et~al.}(2019)\citenamefont {Feng},
  \citenamefont {Jin}, \citenamefont {Tuo},\ and\ \citenamefont
  {Xia}}]{Feng:2018pdq}%
  \BibitemOpen
  \bibfield  {author} {\bibinfo {author} {\bibfnamefont {X.}~\bibnamefont
  {Feng}}, \bibinfo {author} {\bibfnamefont {L.-C.}\ \bibnamefont {Jin}},
  \bibinfo {author} {\bibfnamefont {X.-Y.}\ \bibnamefont {Tuo}}, \ and\
  \bibinfo {author} {\bibfnamefont {S.-C.}\ \bibnamefont {Xia}},\ }\href
  {\doibase 10.1103/PhysRevLett.122.022001} {\bibfield  {journal} {\bibinfo
  {journal} {Phys. Rev. Lett.}\ }\textbf {\bibinfo {volume} {122}},\ \bibinfo
  {pages} {022001} (\bibinfo {year} {2019})},\ \Eprint
  {http://arxiv.org/abs/1809.10511} {arXiv:1809.10511 [hep-lat]} \BibitemShut
  {NoStop}%
\bibitem [{\citenamefont {Detmold}\ and\ \citenamefont
  {Murphy}(2019)}]{Detmold:2018zan}%
  \BibitemOpen
  \bibfield  {author} {\bibinfo {author} {\bibfnamefont {W.}~\bibnamefont
  {Detmold}}\ and\ \bibinfo {author} {\bibfnamefont {D.}~\bibnamefont
  {Murphy}},\ }\href {\doibase 10.22323/1.334.0262} {\bibfield  {journal}
  {\bibinfo  {journal} {PoS}\ }\textbf {\bibinfo {volume} {LATTICE2018}},\
  \bibinfo {pages} {262} (\bibinfo {year} {2019})},\ \Eprint
  {http://arxiv.org/abs/1811.05554} {arXiv:1811.05554 [hep-lat]} \BibitemShut
  {NoStop}%
\bibitem [{\citenamefont {Tuo}\ \emph {et~al.}(2019)\citenamefont {Tuo},
  \citenamefont {Feng},\ and\ \citenamefont {Jin}}]{Tuo:2019bue}%
  \BibitemOpen
  \bibfield  {author} {\bibinfo {author} {\bibfnamefont {X.-Y.}\ \bibnamefont
  {Tuo}}, \bibinfo {author} {\bibfnamefont {X.}~\bibnamefont {Feng}}, \ and\
  \bibinfo {author} {\bibfnamefont {L.-C.}\ \bibnamefont {Jin}},\ }\href
  {\doibase 10.1103/PhysRevD.100.094511} {\bibfield  {journal} {\bibinfo
  {journal} {Phys. Rev. D}\ }\textbf {\bibinfo {volume} {100}},\ \bibinfo
  {pages} {094511} (\bibinfo {year} {2019})},\ \Eprint
  {http://arxiv.org/abs/1909.13525} {arXiv:1909.13525 [hep-lat]} \BibitemShut
  {NoStop}%
\bibitem [{\citenamefont {Detmold}\ and\ \citenamefont
  {Murphy}(2020)}]{Detmold:2020jqv}%
  \BibitemOpen
  \bibfield  {author} {\bibinfo {author} {\bibfnamefont {W.}~\bibnamefont
  {Detmold}}\ and\ \bibinfo {author} {\bibfnamefont {D.}~\bibnamefont {Murphy}}
  (\bibinfo {collaboration} {NPLQCD}),\ }\href@noop {} {\  (\bibinfo {year}
  {2020})},\ \Eprint {http://arxiv.org/abs/2004.07404} {arXiv:2004.07404
  [hep-lat]} \BibitemShut {NoStop}%
\bibitem [{\citenamefont {Tiburzi}\ \emph {et~al.}(2017)\citenamefont
  {Tiburzi}, \citenamefont {Wagman}, \citenamefont {Winter}, \citenamefont
  {Chang}, \citenamefont {Davoudi}, \citenamefont {Detmold}, \citenamefont
  {Orginos}, \citenamefont {Savage},\ and\ \citenamefont
  {Shanahan}}]{Tiburzi:2017iux}%
  \BibitemOpen
  \bibfield  {author} {\bibinfo {author} {\bibfnamefont {B.~C.}\ \bibnamefont
  {Tiburzi}}, \bibinfo {author} {\bibfnamefont {M.~L.}\ \bibnamefont {Wagman}},
  \bibinfo {author} {\bibfnamefont {F.}~\bibnamefont {Winter}}, \bibinfo
  {author} {\bibfnamefont {E.}~\bibnamefont {Chang}}, \bibinfo {author}
  {\bibfnamefont {Z.}~\bibnamefont {Davoudi}}, \bibinfo {author} {\bibfnamefont
  {W.}~\bibnamefont {Detmold}}, \bibinfo {author} {\bibfnamefont
  {K.}~\bibnamefont {Orginos}}, \bibinfo {author} {\bibfnamefont {M.~J.}\
  \bibnamefont {Savage}}, \ and\ \bibinfo {author} {\bibfnamefont {P.~E.}\
  \bibnamefont {Shanahan}},\ }\href {\doibase 10.1103/PhysRevD.96.054505}
  {\bibfield  {journal} {\bibinfo  {journal} {Phys. Rev. D}\ }\textbf {\bibinfo
  {volume} {96}},\ \bibinfo {pages} {054505} (\bibinfo {year} {2017})},\
  \Eprint {http://arxiv.org/abs/1702.02929} {arXiv:1702.02929 [hep-lat]}
  \BibitemShut {NoStop}%
\bibitem [{\citenamefont {Shanahan}\ \emph {et~al.}(2017)\citenamefont
  {Shanahan}, \citenamefont {Tiburzi}, \citenamefont {Wagman}, \citenamefont
  {Winter}, \citenamefont {Chang}, \citenamefont {Davoudi}, \citenamefont
  {Detmold}, \citenamefont {Orginos},\ and\ \citenamefont
  {Savage}}]{Shanahan:2017bgi}%
  \BibitemOpen
  \bibfield  {author} {\bibinfo {author} {\bibfnamefont {P.~E.}\ \bibnamefont
  {Shanahan}}, \bibinfo {author} {\bibfnamefont {B.~C.}\ \bibnamefont
  {Tiburzi}}, \bibinfo {author} {\bibfnamefont {M.~L.}\ \bibnamefont {Wagman}},
  \bibinfo {author} {\bibfnamefont {F.}~\bibnamefont {Winter}}, \bibinfo
  {author} {\bibfnamefont {E.}~\bibnamefont {Chang}}, \bibinfo {author}
  {\bibfnamefont {Z.}~\bibnamefont {Davoudi}}, \bibinfo {author} {\bibfnamefont
  {W.}~\bibnamefont {Detmold}}, \bibinfo {author} {\bibfnamefont
  {K.}~\bibnamefont {Orginos}}, \ and\ \bibinfo {author} {\bibfnamefont
  {M.~J.}\ \bibnamefont {Savage}},\ }\href {\doibase
  10.1103/PhysRevLett.119.062003} {\bibfield  {journal} {\bibinfo  {journal}
  {Phys. Rev. Lett.}\ }\textbf {\bibinfo {volume} {119}},\ \bibinfo {pages}
  {062003} (\bibinfo {year} {2017})},\ \Eprint
  {http://arxiv.org/abs/1701.03456} {arXiv:1701.03456 [hep-lat]} \BibitemShut
  {NoStop}%
\bibitem [{\citenamefont {Feng}\ and\ \citenamefont
  {Jin}(2019)}]{Feng:2018qpx}%
  \BibitemOpen
  \bibfield  {author} {\bibinfo {author} {\bibfnamefont {X.}~\bibnamefont
  {Feng}}\ and\ \bibinfo {author} {\bibfnamefont {L.}~\bibnamefont {Jin}},\
  }\href {\doibase 10.1103/PhysRevD.100.094509} {\bibfield  {journal} {\bibinfo
   {journal} {Phys. Rev. D}\ }\textbf {\bibinfo {volume} {100}},\ \bibinfo
  {pages} {094509} (\bibinfo {year} {2019})},\ \Eprint
  {http://arxiv.org/abs/1812.09817} {arXiv:1812.09817 [hep-lat]} \BibitemShut
  {NoStop}%
\bibitem [{\citenamefont {Ananthanarayan}\ and\ \citenamefont
  {Moussallam}(2004)}]{Ananthanarayan:2004qk}%
  \BibitemOpen
  \bibfield  {author} {\bibinfo {author} {\bibfnamefont {B.}~\bibnamefont
  {Ananthanarayan}}\ and\ \bibinfo {author} {\bibfnamefont {B.}~\bibnamefont
  {Moussallam}},\ }\href {\doibase 10.1088/1126-6708/2004/06/047} {\bibfield
  {journal} {\bibinfo  {journal} {JHEP}\ }\textbf {\bibinfo {volume} {06}},\
  \bibinfo {pages} {047} (\bibinfo {year} {2004})},\ \Eprint
  {http://arxiv.org/abs/hep-ph/0405206} {arXiv:hep-ph/0405206} \BibitemShut
  {NoStop}%
\bibitem [{\citenamefont {Endres}\ \emph {et~al.}(2016)\citenamefont {Endres},
  \citenamefont {Shindler}, \citenamefont {Tiburzi},\ and\ \citenamefont
  {Walker-Loud}}]{Endres:2015gda}%
  \BibitemOpen
  \bibfield  {author} {\bibinfo {author} {\bibfnamefont {M.~G.}\ \bibnamefont
  {Endres}}, \bibinfo {author} {\bibfnamefont {A.}~\bibnamefont {Shindler}},
  \bibinfo {author} {\bibfnamefont {B.~C.}\ \bibnamefont {Tiburzi}}, \ and\
  \bibinfo {author} {\bibfnamefont {A.}~\bibnamefont {Walker-Loud}},\ }\href
  {\doibase 10.1103/PhysRevLett.117.072002} {\bibfield  {journal} {\bibinfo
  {journal} {Phys. Rev. Lett.}\ }\textbf {\bibinfo {volume} {117}},\ \bibinfo
  {pages} {072002} (\bibinfo {year} {2016})},\ \Eprint
  {http://arxiv.org/abs/1507.08916} {arXiv:1507.08916 [hep-lat]} \BibitemShut
  {NoStop}%
\end{thebibliography}%

\end{document}